\documentclass[12pt,a4paper,english,aps,manuscript]{revtex4}
\usepackage[T1]{fontenc}
\usepackage[latin9]{inputenc}
\usepackage{rotfloat}
\usepackage{dvipost}
\usepackage{textcomp}
\usepackage{amsbsy}
\usepackage{amstext}
\usepackage{graphicx}
\usepackage{esint}
\usepackage{subscript}

\makeatletter

\dvipostlayout
\dvipost{osstart color push Red}
\dvipost{osend color pop}
\dvipost{cbstart color push Blue}
\dvipost{cbend color pop}
\DeclareRobustCommand{\lyxadded}[3]{\changestart#3\changeend}
\DeclareRobustCommand{\lyxdeleted}[3]{%
\changestart\overstrikeon#3\overstrikeoff\changeend}

\@ifundefined{textcolor}{}
{%
 \definecolor{BLACK}{gray}{0}
 \definecolor{WHITE}{gray}{1}
 \definecolor{RED}{rgb}{1,0,0}
 \definecolor{GREEN}{rgb}{0,1,0}
 \definecolor{BLUE}{rgb}{0,0,1}
 \definecolor{CYAN}{cmyk}{1,0,0,0}
 \definecolor{MAGENTA}{cmyk}{0,1,0,0}
 \definecolor{YELLOW}{cmyk}{0,0,1,0}
}

\newcommand{\Htwop}{\hbox{\rm H}_2^+}
\newcommand{\Dtwop}{\hbox{\rm D}_2^+}
\newcommand{\Ttwop}{\hbox{\rm T}_2^+}
\newcommand{\HDp}{\hbox{\rm HD}^+}
\newcommand{\HTp}{\hbox{\rm HT}^+}

\makeatother

\usepackage{babel}
\begin{document}

\title{The electric quadrupole moment of molecular hydrogen ions and their
potential for a molecular ion clock }

\author{D. Bakalov }

\affiliation{Institute for Nuclear Research and Nuclear Energy, Tsarigradsko chaussée
72, Sofia 1784, Bulgaria }

\author{S. Schiller}

\affiliation{Institut für Experimentalphysik, Heinrich-Heine-Universität Düsseldorf,
40225 Düsseldorf, Germany}
\begin{abstract}
The systematic shifts of the transition frequencies in the
molecular hydrogen ions are of relevance to ultra-high-resolution
radio-frequency, microwave and optical spectroscopy of these
systems, performed in ion traps. We develop the ab-initio
description of the interaction of the electric quadrupole moment
of this class of molecules with the static electric field
gradients present in ion traps. In good approximation, it is
described in terms of an effective perturbation hamiltonian. An
approximate treatment is then performed in the Born-Oppenheimer
approximation. We give an expression of the electric quadrupole
coupling parameter valid for all hydrogen molecular ion species
and evaluate it for a large number of states of $\Htwop\,$,
$\HDp$, and $\Dtwop$. The systematic shifts can be evaluated as
simple expectation values of the perturbation hamiltonian. Results
on radio-frequency (M1), one-photon electric dipole (E1) and
two-photon E1 transitions between hyperfine states in $\HDp$ are
reported. For two-photon E1 transitions between rotationless
states the shifts vanish. For a large subset of rovibrational
one-photon transitions the absolute values of the quadrupole
shifts range from 0.3 to 10~Hz for an electric field gradient of
10\textsuperscript{8} V/m\textsuperscript{2}. We point out an
experimental procedure for determining the quadrupole shift which
will allow reducing its contribution to the uncertainty of
unperturbed rovibrational transition frequencies to the
$1\times10$\textsuperscript{-15} relative level and, for selected
transitions, even below it. The combined contributions of
black-body radiation, Zeeman, Stark and quadrupole effects are
considered for a large set of transitions and it is estimated that
the total transition frequency uncertainty of selected transitions
can be reduced below the $1\times10^{-15}$ level.
\end{abstract}
\maketitle

\section{Introduction}

One of the fascinating aspects of the ion trap invented by W. Paul
and its later variants is the suitability for trapping a wide
variety of particles. While atomic ions are the most
\lyxadded{schiller}{Sat Oct 26 23:23:13 2013}{frequently }studied
particle types, today, cold molecular ions are being studied in an
increasing number of laboratories world-wide. The molecular ion
most intensely studied \lyxadded{schiller}{Sat Oct 26 23:23:25
2013}{so far }from a spect\lyxadded{schiller}{Sat Oct 26 23:23:37
2013}{r}oscopic point of view is the molecular hydrogen ion
$\HDp$, for which significant progress has been made in the last
decade, both on the experimental \cite{SS3,Bressel 2012} and on
the ab-initio theory front (see Ref.~\cite{Korobov 2012} and
references therein). Combined studies of $\HDp$ and of the
isotopologue molecules ($\Htwop$ \cite{Karr 2011}, $\HTp$
\cite{Bekbaev 2011}, $\Dtwop$, etc.), may in the near future lead
to the determination of several fundamental physical constants,
such as the ratios of proton, deuteron and triton mass relative to
the electron mass, and the Rydberg energy,
etc.\cite{SS1,SS2,SS3,mkajita} with potentially competitive
accuracy and with a different experimental approach than in atomic
laser spectroscopy and Penning trap spectroscopy. A first step in
this direction has been performed with two laser-spectroscopic
measurements on $\HDp$ \cite{SS3,Bressel 2012}, from which the
ratio of the electron mass to the reduced nuclear mass can be
inferred with a relative experimental inaccuracy of approximately
4 and 2 parts in $10^{9}$, respectively.

Moreover, the molecular hydrogen ions may be suited to investigate
the question whether the mentioned dimensionless fundamental constants
are independent of time \cite{SS2} and of location in space, a postulate
made by the principle of local position invariance of General Relativity.

These possibilities are only feasible if the experimental
uncertainty in the measurement of transition frequencies can be
reduced to a level necessary for the particular application. For
example, in order to make competitive determinations of the
fundamental constants, (currently) uncertainties of
$1\times10^{-10}$ or less are desirable, while for the
investigation of their time-independence, $1\times10^{-16}$
\lyxadded{schiller}{Sat Oct 26 23:25:14 2013}{or less }is
desirable. A series of systematic effects needs to be carefully
taken into account, including the effects of the external electric
and magnetic fields in the volume occupied by the molecular ions.
The Zeeman shift of the transition frequency induced by the weak
magnetic fields usually present in experiments was thoroughly
investigated in \cite{Zee1,Zee2}. Various aspects of the Stark
effect of the $\HDp$ molecule have been studied in
\cite{moss,older}, and recently in \cite{kool,hfi}.

In the present paper we determine theoretically the energy shifts
caused by the interaction of the permanent electric quadrupole
moment of the molecular ion with the inhomogeneities of the
electric field of the ion trap. For atomic ions used in optical
clocks, this is a well-known systematic effect, but for molecular
ions, this effect has not been treated before for any molecule, to
the best of our knowledge.

Concerning related work, we mention that the electric quadrupole
transitions of the molecular hydrogen ions have been of some
theoretical interest. The transition matrix elements for $\Htwop$
have first been treated by Bates and Poots \cite{Bates and Poots}
and later more extensively in Refs. \cite{Peek,Posen Dalgarno
Peek,Pilon and Baye}; the value of the permanent quadrupole moment
in the vibrational ground state $v=0$ is reported in
Refs.~\cite{Bates and Poots,Peek,Bishop and Lam 1987,Bishop and
Lam 1988}. To our knowledge, there is only a single calculation
concerning $\HDp$, namely of its permanent quadrupole moment in
the level $v=0$, in Ref.~\cite{Bathia}. Recently, the quadrupole
transition moments for $\Dtwop$ have been reported \cite{Pilon}.

After developing the general theory in Sec.~II, as in a previous
work \cite{hfi} the numerical calculations are performed in the
Born-Oppenheimer approach, introduced in Sec. III, which provides
rovibrational energy levels and matrix elements with relative
accuracy of approximately $10^{-3}$, but is entirely sufficient
for the evaluation of the electric quadrupole effect in ion traps.
This will be justified \textit{a posteriori} by the small size of
the calculated corrections. The detailed study of a large number
of transitions in $\HDp$ is given in Sec. IV. The discussion
(Sec.~V) shows that Zeeman, electric quadrupole, and Stark shifts
can be controlled to a sufficient level even in spectroscopy
aiming for\lyxdeleted{Schiller}{Thu Sep 12 21:20:39 2013}{ } high
accuracy.

\section{Electric quadrupole shift in three-particle bound systems}

In this section we derive the general expressions for the quadrupole
interaction effect in a three-body bound system.

We use the Jacobi coordinate vectors of the three-body system, $\mathbf{R}_{C}$,
$\mathbf{R}$ and $\mathbf{r}$, which are related to the individual
particle position vectors $\mathbf{R}_{k},\ k=1,\,2,\,3$ by means
of
\begin{eqnarray}
\mathbf{R}_{C} & = & \sum\limits _{k=1}^{3}\frac{m_{k}}{m_{t}}\mathbf{R}_{k}\,,\nonumber \\
\mathbf{R} & = & \mathbf{R}_{2}-\mathbf{R}_{1}\,,\nonumber \\
\mathbf{r} & = & \mathbf{R}_{3}-\frac{m_{1}}{m_{12}}\mathbf{R}_{1}-\frac{m_{2}}{m_{12}}\mathbf{R}_{2}\,,\nonumber \\
m_{kk'} & = & m_{k}+m_{k'}\,,\nonumber \\
m_{k,k'} & = & \frac{m_{k}m{}_{k'}}{m_{kk'}}\,,\nonumber \\
m_{t} & = & \sum_{k}m_{k}\,,
\end{eqnarray}
where $m_{k}$ are the masses of the particles. In the $\HDp$ ion,
$k=1,\,2,\,3$ labels the deuteron, the proton and the electron, respectively.
Note that $\mathbf{r}$ is defined as the radius vector of the electron
reckoned from the center of mass of the two nuclei. In terms of the
Jacobi vectors, the non-relativistic Hamiltonian $H_{NR}$ splits
into the sum of the free Hamiltonian $H_{C}$ of the system ``as a
whole'' and the Hamiltonian $H$ of the internal degrees of freedom:
\begin{equation}
H_{NR}=H_{C}+H,\ \,\,\,\,\,\,\, H_{C}=\frac{\mathbf{P}_{C}^{2}}{2m_{t}}\,,\label{separ0n}
\end{equation}

\begin{equation}
H=\frac{\mathbf{P}^{2}}{2m_{1,2}}+\frac{\mathbf{p}^{2}}{2m_{3,12}}+V(\mathbf{R},\mathbf{r})\,,
\end{equation}

\begin{equation}
V(\mathbf{R},\mathbf{r})=\sum\limits _{k<k'}\frac{Z_{k}Z_{k'}e^{2}}{|\mathbf{R}_{k}-\mathbf{R}_{k'}|}\,,
\end{equation}
where $\mathbf{P}_{C}$, $\mathbf{P}$ and $\mathbf{p}$ are the momenta
conjugate to $\mathbf{R}_{C}$, $\mathbf{R}$ and $\mathbf{r}$, respectively,
and $Z_{k}$ are the particle charges in units of $e$.

In an external electric potential $U$, the non-relativistic
Hamiltonian $H_{NR}$ acquires an additional term:
$H_{NR,ext}=H_{NR}+\Delta H$ with
\begin{equation}
\ensuremath{\Delta H=\sum\limits _{k=1}^{3}eZ_{k}U(\mathbf{R}_{k})}
\end{equation}
being the electrostatic energy of the particles. For external fields
that vary slowly in space and time, $\Delta H$ is approximated with
the truncated multipole expansion

\begin{eqnarray}
\Delta H & = & \Delta H_{0}+\Delta H_{d}+\Delta H_{Q\,},\nonumber \\
\Delta H_{0} & = & (e\sum_{k}Z_{k})U(\mathbf{R}_{C})\,,\nonumber \\
\Delta H_{d} & = & -\mathbf{d}_{C}\cdot\mathbf{E}(\mathbf{R}_{C})\,,\label{eq:expan0}\\
\Delta H_{Q} & = & -\frac{1}{3}\Theta_{C}\cdot Q(\mathbf{R}_{C})\,,\nonumber
\end{eqnarray}
where $\mathbf{d}_{C}$ is the electric dipole moment of the system
with respect to $\mathbf{R}_{C}$, $\mathbf{d}_{C}=\sum\limits _{k}eZ_{k}\mathbf{r}_{k}\,,$
$\mathbf{r}_{k}=\mathbf{R}_{k}-\mathbf{R}_{C}$~, $\Theta_{C}$ is
the irreducible tensor of rank 2 of the quadrupole moment with Cartesian
components

\begin{equation}
(\Theta_{C})_{ij}=(3/2)\sum_{k}eZ_{k}(r_{ki}r_{kj}-\delta_{ij}\mathbf{r}_{k}^{2}/3),
\end{equation}

\begin{equation}
\ensuremath{\mathbf{E}(\mathbf{R}_{C})=-\nabla U(\mathbf{x})\vert_{\mathbf{x}=\mathbf{R}_{C}}}
\end{equation}
is the external electric field at the center point, and
\begin{equation}
\ensuremath{Q(\mathbf{R}_{C})_{ij}=-(\partial^{2}/\partial x_{i}\,\partial x_{j})U(\mathbf{x})\vert_{\mathbf{x}=\mathbf{R}_{C}}}\,.
\end{equation}
$\Delta H_{0}$, together with $H_{C}$, are dropped because of being
related to the degrees of freedom of the 3-body system ``as a whole''.
Different aspects of the second-order perturbation contribution of
the dipole term have been evaluated in \cite{moss,kool,hfi}. In what
follows, we focus our attention on the contribution of the quadrupole
interaction term $\Delta H_{Q}$ in first order of perturbation theory.

The Cartesian components $(\Theta_{C})_{ij}$ in terms of the Cartesian
components of the vectors $\mathbf{R}$, and $\mathbf{r}$ (in the
center-of-mass frame $\mathbf{R}_{C}=0$) are

\begin{equation}
(\Theta_{C})_{ij}=\frac{3}{2}e\left(a_{0}\left(R_{i}R_{j}-\frac{\delta_{ij}}{3}\mathbf{R}^{2}\right)+a_{1}\left(\frac{R_{i}r_{j}+r_{i}R_{j}}{2}-\frac{\delta_{ij}}{3}\mathbf{R}\cdot\mathbf{r}\right)-a_{2}\left(r_{i}r_{j}-\frac{\delta_{ij}}{3}\mathbf{r}^{2}\right)\right),\label{eq:thetacm-cart}
\end{equation}

\begin{equation}
a_{0}=(m_{1}^{2}+m_{2}^{2})/m_{12}^{2},\ a_{1}=2(m_{2}-m_{1})m_{3}/(m_{12}m_{t}),\ a_{2}=(m_{12}^{2}-2m_{3}^{2})/m_{t}^{2}\,.\label{eq:as}
\end{equation}
Note the factor $3/2$ in the definition of $\Theta_{C}$ that is
not present in the analogous expressions in Refs.~\cite{Bates and
Poots,Bathia}. In evaluating the matrix elements of $\Theta_{C}$
in the angular momentum representation, similar to
Refs.~\cite{pra57,prl79} we use the expansion of the
non-relativistic three-body wave function of the bound state with
the orbital momentum quantum number $L$, the projection of
$\mathbf{L}$ on the space-fixed $z$-axis equal to $M$, the
vibrational quantum number $v$ and the parity $\lambda$ in the
basis of the symmetrized Wigner functions ${\cal D}_{Mm}^{\lambda
L}$,

\begin{equation}
\psi^{\lambda vLM}(\mathbf{R},\mathbf{r})=\langle\mathbf{R},\mathbf{r}|\lambda vLM\rangle=\sum\limits _{m=0}^{L}u_{m}^{\lambda vL}(R,r,\gamma){\cal D}_{Mm}^{\lambda L}(\Phi,\theta,\varphi),\label{eq:wigner-exp}
\end{equation}
where $\gamma$ is the angle between the vectors $\mathbf{R}$ and
$\mathbf{r}$: $\cos\gamma=\mathbf{R}\cdot\mathbf{r}/(Rr)$, while
$\Phi$, $\theta$ and $\varphi$ are the Euler angles of the rotation
that transforms the space-fixed into the body-fixed reference frame
with $z$-axis along $\mathbf{R}$ and $\mathbf{r}$ in the $xOz$
plane.

The amplitudes $u_{m}^{\lambda vL}(R,r,\gamma)$ are normalized by
the condition $\int dR\, R^{2}\int dr\, r^{2}\int d\gamma\,\sin\gamma\sum_{m}(u_{m}^{\lambda vL}(R,\, r,\,\gamma))^{2}=1$.

The normalized symmetrized Wigner functions ${\cal D}_{Mm}^{\lambda L}(\Phi,\theta,\varphi)$
are linear combinations with definite parity of the complex conjugated
standard Wigner functions:

\begin{equation}
{\cal D}_{Mm}^{\lambda L}(\Phi,\theta,\varphi)=\sqrt{\frac{2L+1}{16\pi^{2}(1+\delta_{0m})}}\left((-1)^{m}D_{Mm}^{L*}(\Phi,\theta,\varphi)+\lambda(-1)^{L}D_{M-m}^{L*}(\Phi,\theta,\varphi)\right)\,.\label{eq:wigner}
\end{equation}
Next, the cyclic components of the quadrupole moment $\tilde{\Theta}_{C}$
(labeled with the tilde to distinguish from the Cartesian components)
are put in the form of a sum of terms with factorized dependence on
the sets of angular and radial variables:
\begin{eqnarray}
\tilde{\Theta}_{C} & = & e\bigg(a_{0}R^{2}X^{0}+a_{1}\, R\, r\,\big(d_{00}^{1}(\gamma)X^{0}-\frac{\sqrt{3}}{2}d_{10}^{1}(\gamma)X^{1}\big)\nonumber \\
 &  & -a_{2}\, r^{2}\big(d_{00}^{2}(\gamma)X^{0}+d_{10}^{2}(\gamma)X^{1}+d_{20}^{2}(\gamma)X^{2}\big)\bigg)\,,\label{eq:theta_c}
\end{eqnarray}
where $d_{mM}^{L}(\gamma)$ are the ``small'' Wigner $d$-matrices
given in \cite{Edmonds}. The $(X^{i})_{0},\, i=0,\,1,\,2$ are the
zero-th cyclic components of irreducible tensor operators $X^{i}$
of rank 2 acting on the angular variables:

\begin{equation}
(X^{0})_{0}=\frac{3}{2}\cos^{2}\theta-\frac{1}{2},\ \,\,(X^{1})_{0}=\sqrt{6}\sin\theta\cos\theta\cos\varphi,\ \,\,(X^{2})_{0}=\sqrt{\frac{3}{2}}\sin^{2}\theta\cos2\varphi.
\end{equation}
The reduced matrix elements of $X^{i}$ in the angular basis of Eq.~(\ref{eq:wigner})
have the form:
\begin{eqnarray}
\langle\lambda'm'L'||X^{0}||\lambda mL\rangle & = & N(C_{Lm,20}^{L'm'}+\sigma C_{L-m,20}^{L'm'})\,,\nonumber \\
\langle\lambda'm'L'||X^{1}||\lambda mL\rangle & = & N(C_{Lm,2-1}^{L'm'}-C_{Lm,21}^{L'm'}+\sigma(C_{L-m,2-1}^{L'm'}-C_{L-m,21}^{L'm'}))\,,\label{eq:explicitX}\\
\langle\lambda'm'L'||X^{2}||\lambda mL\rangle & = & N(C_{Lm,2-2}^{L'm'}+C_{Lm,22}^{L'm'}+\sigma(C_{L-m,2-2}^{L'm'}+C_{L-m,22}^{L'm'}))\,,\nonumber
\end{eqnarray}
where $N=\delta_{\lambda\lambda'}\sqrt{(2L+1)/((1+\delta_{0m})(1+\delta_{0m'}))},$~$\sigma=\lambda(-1)^{m+L}$,
and $C_{a\alpha,b\beta}^{e\varepsilon}\equiv\langle e\varepsilon|a\alpha,b\beta\rangle$
are the Clebsch-Gordan coefficients.

Thus, the matrix elements of $\Delta H_{Q}$ in the basis Eq.~(\ref{eq:wigner-exp})
become

\begin{equation}
\langle\lambda v'L'M'|\Delta H_{Q}|\lambda vLM\rangle=-\frac{1}{3}\bigg(\sum\limits _{q=-2}^{2}\tilde{Q}^{q}(\mathbf{R}_{C})\, C_{LM,2q}^{L'M'}\bigg)\,(2L'+1)^{-1/2}\langle\lambda v'L'||\tilde{\Theta}_{C}||\lambda vL\rangle,
\end{equation}

\[
\langle\lambda v'L'||\tilde{\Theta}_{C}||\lambda vL\rangle=e\sum\limits _{m'm}\bigg(\langle\lambda m'L'||X^{0}||\lambda mL\rangle\left(a_{0}I_{\lambda,v'L',vL}^{(00)m'm}+a_{1}I_{\lambda,v'L',vL}^{(01)m'm}-a_{2}I_{\lambda,v'L',vL}^{(02)m'm}\right)
\]
\begin{equation}
-\langle\lambda m'L'||X^{1}||\lambda mL\rangle\left(\frac{\sqrt{3}}{2}a_{1}I_{\lambda,v'L',vL}^{(11)m'm}+a_{2}I_{\lambda,v'L',vL}^{(12)m'm}\right)-\langle\lambda m'L'||X^{2}||\lambda mL\rangle a_{2}I_{\lambda,v'L',vL}^{(22)m'm}
\label{eq:matrel-gen-sh}
\end{equation}
where $\tilde{Q^{q}}$ are the contravariant cyclic components of
$Q$ and $I_{\lambda,v'L',vL}^{(kn)m'm}$ denote the following integrals,

\begin{equation}
I_{\lambda,v'L',vL}^{(kn)m'm}=\int dR\, R^{2}\int dr\, r^{2}\int d\gamma\sin(\gamma)\, u_{m'}^{\lambda v'L'}(R,r,\gamma)\, R^{2-n}r^{n}\, d_{k0}^{n}(\gamma)u_{m}^{\lambda vL}(R,r,\gamma)\,.\label{eq:integrals}
\end{equation}
Eqs.~(\ref{eq:theta_c}-\ref{eq:matrel-gen-sh}), after the appropriate
changes of variables in Eqs.~(\ref{eq:theta_c}) and (\ref{eq:integrals})
can be used with any alternative choice of the arguments of the radial
amplitudes $u$ in the expansion Eq.~(\ref{eq:wigner-exp}), e.g.
the variables $|\mathbf{R}_{k}-\mathbf{R}_{k'}|,\, k<k'$ or their
linear combinations \cite{prl79}, but need be reworked for alternative
basis sets in the space of functions of the angular variables, such
as the expansion in bi-harmonics of Refs.~\cite{SS1,pra59}.

The quadrupole term $\Delta H_{Q}=-(1/3)\Theta_{C}\cdot
Q(\mathbf{R}_{C})$ in the expansion Eq.~(\ref{eq:expan0}) couples,
in the general case, states with different values of the orbital
momentum $L$ and its projection $M$ and shifts the energy levels
of the three-body states by amounts that depend on $M$.

Previous studies of the effects of external magnetic fields
\cite{Zee2} had demonstrated the advantages of considering the
various perturbations to the dominating Coulomb interactions due
to relativistic effects, particle spin and external fields on the
same footing. An efficient implementation of these calculations in
first order of perturbation theory is the use of an ``effective
Hamiltonian'' $H_{{\rm eff}}$. We remind that the ``effective spin
Hamiltonian'' of an atomic system is the projection of the spin
interaction operator on the finite dimensional space of
eigenstates of the non-relativistic Hamiltonian of the system with
definite values of the orbital angular momentum and the remaining
non-relativistic quantum numbers, in which couplings to different
$L$ are neglected.

We therefore include the effects  of the quadrupole interaction $\Delta H_{Q}$
in the form of an additional term $V^{\mathrm{Q}}$ in the effective
spin Hamiltonian $H_{{\rm eff}}^{{\rm hfs}}$, introduced in \cite{prl06}
(denoted by $H_{{\rm eff}}$ there) in the calculation of the hyperfine
structure and completed to $H_{{\rm eff}}^{{\rm tot}}=H_{{\rm eff}}^{{\rm hfs}}+V^{{\rm mag}}$
by terms $V^{\mathrm{mag}}$ that describe the Zeeman shifts in \cite{Zee2}.
That is, we set

\begin{eqnarray}
V^{\mathrm{Q}}(v,\, L) & = & E_{14}(v,\, L)\, Q(\mathbf{R}_{C})\cdot(\mathbf{L}\otimes\mathbf{L})^{(2)},\nonumber \\
H_{{\rm eff}}^{{\rm tot+Q}}(v,\, L) & = & H_{{\rm eff}}^{{\rm hfs}}(v,\, L)+V^{{\rm mag}}(v,\, L)+V^{\mathrm{Q}}(v,\, L)\,,\label{eq:heff}
\end{eqnarray}
where $(\mathbf{L}\otimes\mathbf{L})^{(2)}$ is the tensor square
of the orbital momentum operator $\mathbf{L}$ - the only irreducible
tensor operator of rank 2 acting in the space of states with definite
value of $L$. In Eq.(\ref{eq:heff}) we have shown explicitly the
dependence of the effective Hamiltonian and its various terms on the
quantum numbers $(v,\, L)$ of the non-relativistic state to which
they refer. From the next section on, in order to simplify the notations
we shall omit these quantum numbers while keeping in mind the dependence
on them. The advantage of using the effective Hamiltonian is that
the integrals of the 3-body wave functions of Eqs.~(\ref{eq:twoc-trunc})
or (\ref{eq:wigner-exp}) over $R$, $r$ and $\gamma$ are encoded
in the single constant $E_{14}$, so that the electric quadrupole
shift of each individual quantum state is calculated by standard angular
momentum algebra.

The expression for $E_{14}$ reads:

\begin{eqnarray}
E_{14}(v,\, L) & = & -\frac{1}{3}\frac{\langle\lambda vL||\tilde{\Theta}_{C}||\lambda vL\rangle}{\langle L||(\mathbf{L}\otimes\mathbf{L})^{(2)}||L\rangle}\,,\label{eq:E14g}\\
\langle L||(\mathbf{L}\otimes\mathbf{L})^{(2)}||L\rangle & = & \sqrt{\frac{\Gamma(2L+4)}{4!\Gamma(2L-1)}}\,.\nonumber
\end{eqnarray}

\section{Born-Oppenheimer approximation}

The gradient of the electric field acting on an ion in a
quadrupole ion trap is of the order of
$10^{8}$~V/m\textsuperscript{2} and in what follows it will be
shown that this magnitude gives rise to energy level shifts not
exceeding 100~Hz, significantly below the Zeeman shifts of most
levels for the typical fields that are applied in ion traps
\cite{Zee1,Zee2}. This situation softens the requirements to the
numerical and theoretical accuracy of the treatment, and allows
for using the Born-Oppenheimer wave functions instead of the
highly accurate variational wave functions of Ref. \cite{kor-HD}.

The Born-Oppenheimer approximation assumes that instead of $\mathbf{R}_{C}$
the molecular ion's ``motion as a whole'' is associated with the nuclear
center-of-mass position vector $\mathbf{R}_{B}=(m_{1}\mathbf{R}_{1}+m_{2}\mathbf{R}_{2})/m_{12}$
and its conjugate momentum $\mathbf{P}_{B}$. $H_{NR}$ then takes
the form

\begin{eqnarray}
H_{NR} & = & H_{B}+\Delta H_{B}+H\,,\nonumber \\
H_{B} & = & \frac{\mathbf{P}_{B}^{2}}{2\, m_{12}}\,,\nonumber \\
\Delta H_{B} & = & \frac{2}{m_{12}}(\mathbf{P}_{B}\cdot\mathbf{p})\,,
\end{eqnarray}
where $H$ is that part that depends only on the internal degrees
of freedom. Separation of external and internal degrees of freedom
occurs by neglecting the cross term $\Delta H_{B}$. This neglect
limits \textit{a priori} the relative accuracy of the results to the
magnitude of the omitted terms, of order $O(4\, m_{3}/m_{12})\sim10^{-3}$.
The inaccuracy due to the replacement of $m_{t}$ in the denominator
of $H_{C}$ in Eq.~(\ref{separ0n}) by $m_{12}$ in $H_{B}$ is smaller.

In order to further separate the degrees of freedom of the electron
from the relative motion of the nuclei we expand the wave function
of the eigenstates of $H$ in the basis of eigenfunctions of the electronic
Hamiltonian:

\begin{equation}
\psi^{\lambda vLM}(\mathbf{R},\mathbf{r})=\sum\limits _{c}\psi_{c}^{({\rm N)}\lambda vLM}(\mathbf{R})\psi_{c}^{({\rm e)}}(\mathbf{r};R)\ ,
\end{equation}

\begin{equation}
(H^{({\rm e)}}-E_{c}(R))\psi_{c}^{({\rm e)}}(\mathbf{r};R)=0\ ,\label{eq:twoc}
\end{equation}

\begin{equation}
H=H^{({\rm N)}}+H^{({\rm e)}},\ \ H^{({\rm N)}}=\frac{1}{2\, m_{1,2}}\mathbf{P}^{2}+\frac{e^{2}}{R}\ ,\ \ H^{({\rm e)}}=\frac{1}{2\, m_{3,12}}\mathbf{p}^{2}-\sum\limits _{k=1,2}\frac{e^{2}}{|\mathbf{R}_{3}-\mathbf{R}_{k}|}.
\end{equation}

We solved Eq.(\ref{eq:twoc}) numerically using its separability in
the prolate spheroidal coordinates

\begin{equation}
\xi=\frac{1}{R}(|\mathbf{R}_{3}-\mathbf{R}_{1}|+|\mathbf{R}_{3}-\mathbf{R}_{2}|),\ \eta=\frac{1}{R}(|\mathbf{R}_{3}-\mathbf{R}_{1}|-|\mathbf{R}_{3}-\mathbf{R}_{2}|)\ ,
\end{equation}

Their definition ranges are $1\leq\xi<\infty,$ $-1<\eta<1$. These
coordinates are related to $r$ and $\gamma$ of Eq.~(\ref{eq:wigner-exp})
by means of

\begin{equation}
r=R\,\sqrt{\frac{m_{1}}{m_{12}}\left(\frac{\xi+\eta}{2}\right)^{2}+\frac{m_{2}}{m_{12}}\left(\frac{\xi-\eta}{2}\right)^{2}-\frac{m_{1}m_{2}}{m_{12}^{2}}},\ \ \cos\gamma=\frac{R}{2r}\left(\xi\eta+\frac{m_{1}-m_{2}}{m_{12}}\right)\ .
\end{equation}

We reproduced the results for $E_{k}(R)$ of Ref.~\cite{twocent}.

The calculations of the dipole polarizabilities of the lower ro-vibrational
states of $\HDp$ in Ref.~\cite{hfi} have shown (by comparison with
the high precision variational results of Ref.~\cite{kor-dip}) that
a relative accuracy of $10^{-3}$ in the computation of the energy
values and of the dipole moments may be reached by keeping only the
first term $c=(1s\sigma)$ in the expansion Eq.(\ref{eq:twoc}) and
by neglecting the diagonal correction term $\langle\psi_{1s\sigma}^{({\rm e)}}|\mathbf{P}^{2}|\psi_{1s\sigma}^{({\rm e)}}\rangle$.
We therefore adopted this approximation in the evaluation of the quadrupole
shift as well and took the wave functions of ``normal'' parity $\lambda=+1$
(the index $\lambda$ is omitted in what follows) in the form:

\begin{equation}
\psi^{vLM}(\mathbf{R},\,\mathbf{r})=R^{-1}\chi_{1s\sigma}^{vL}(R)\,\mathrm{Y}_{LM}(\Phi,\theta)\,\psi_{1s\sigma}^{{\rm (e)}}(\xi,\eta;R)\ ,\label{eq:twoc-trunc}
\end{equation}
with normalization conditions
\begin{equation}
\frac{R^{3}}{8}\ensuremath{\int\int d\xi\, d\eta\,(\xi^{2}-\eta^{2})\,\psi_{1s\sigma}^{{\rm (e)}}(\xi,\eta;R)^{2}=1\ ,}\label{eq:normalization xi eta}
\end{equation}

\begin{equation}
\int_{0}^{^{\infty}}dR\,\chi_{1s\sigma}^{vL}(R)^{2}=1\ .
\end{equation}
We calculated numerically the $\chi_{1s\sigma}^{vL}(R)$ as solutions
of the radial Schrödinger equation.

One could then obtain $E_{14}$ by using Eq.~(\ref{eq:E14g}) and
evaluating the integrals in Eq.~(\ref{eq:integrals}) with the wave
functions of Eq.~(\ref{eq:twoc-trunc}). Instead, we re-expand $\Delta H$
of Eq.~(\ref{eq:expan0}) around the ``Born-Oppenheimer central point''
$\mathbf{R}_{B}$ so that the quadrupole interaction term takes the
form

\begin{equation}
\Delta H_{Q}=-(1/3)\,\Theta_{B}\cdot Q(\mathbf{R}_{B}).\label{eq:delta_HQ_BO}
\end{equation}
The tensor $\Theta_{B}$ differs from $\Theta_{C}$ of Eq.~(\ref{eq:thetacm-cart})
by terms of order $O(a_{1})\sim10^{-4}$ or smaller:

\begin{equation}
(\Theta_{B})_{ij}=e\,\frac{3}{2}\,\left(a_{0}\left(R_{i}R_{j}-\frac{\delta_{ij}}{3}\mathbf{\mathbf{R}}^{2}\right)-\left(r_{i}r_{j}-\frac{\delta_{ij}}{3}\mathbf{r}^{2}\right)\right)\ .\label{eq:theta_B}
\end{equation}
and the error due to replacing $\Theta_{C}$ by $\Theta_{B}$ is within
the adopted accuracy limits. Note that $a_{0}=\frac{1}{2}$ for the
homonuclear ions $\Htwop$, $\Dtwop$, $\Ttwop$ but differs for the
heteronuclear ones.

Similar to Eq.~(\ref{eq:theta_c}), we expand the cyclic components
$\tilde{\Theta}_{B}$ over the set of irreducible tensor operators
$X^{i}$, but keep only the terms involving $X^{0}$ since the matrix
elements of $X^{i},\, i\ge1$ vanish in the $\sigma$-term approximation
with $m'=m=0$, adopted in Eq.~(\ref{eq:twoc-trunc}):

\begin{equation}
\ensuremath{\tilde{\Theta}_{B}\approx e\left(a_{0}R^{2}-r^{2}d_{00}^{2}(\gamma)\right)X^{0}}.
\end{equation}
In order to facilitate comparison with the results of earlier papers
on the subject, instead of using the more general notations of Eq.~(\ref{eq:integrals}),
we put the reduced matrix elements of $\tilde{\Theta}_{B}$ in the
form:

\begin{eqnarray}
\langle\lambda v'L'||\tilde{\Theta}_{B}||\lambda vL\rangle & \approx & \langle\lambda0L'||X^{0}||\lambda0L\rangle\,\overline{M}_{v'L',vL}\,,\nonumber \\
\overline{M}_{v'L',vL} & = & e\,\int_{0}^{^{\infty}}dR\,\chi_{1s\sigma}^{v'L'}(R)\, M(R)\,\chi_{1s\sigma}^{vL}(R)\,,\nonumber \\
M(R) & = & R^{2}\left(\frac{1}{2}-\frac{m_{1}m_{2}}{m_{12}^{2}}\right)+F(R)\,,\nonumber \\
F(R) & = & R^{2}\bigg(\frac{1}{2}+\frac{R^{3}}{8}\int d\xi\, d\eta\,(\xi^{2}-\eta^{2})
 \times\nonumber \\
 &  & \times\frac{1}{8}(\xi^{2}+\eta^{2}-3-3\xi^{2}\eta^{2})\left(\psi_{1s\sigma}^{(e)}
 (\xi,\eta;R)\right)^{2}\bigg)\ .\label{eq:thitaBred}
\end{eqnarray}

We have made use of the symmetry of the wavefunction squared $\psi_{1s\sigma}^{(e)}{}^{2}$
with respect to $\eta\rightarrow-\eta$, so that only terms with even
powers of $\eta$ contribute.

The function $M(R)$ may also be expressed as $M(R)=R^{2}(m_{1}^{2}+m_{2}^{2})/m_{12}^{2}+\langle z^{2}\rangle+\langle x^{2}\rangle$
(the angular brackets refer to the averaging over the electronic coordinates
with $\psi_{1s\sigma}^{({\rm e)}}$). Note that $F(R)$ is independent
of the molecular species and $M(R)$ is the same for all homonuclear
species: $M(R)|_{m_{1}=m_{2}}=R^{2}/4+F(R)$. The function $F(R)$
, which gives the correction to the asymptotic behavior of $M(R)$,
was introduced in \cite{Peek}. In Fig. \ref{fig:function N} we plot
it.


\begin{figure}
\centering{}\includegraphics{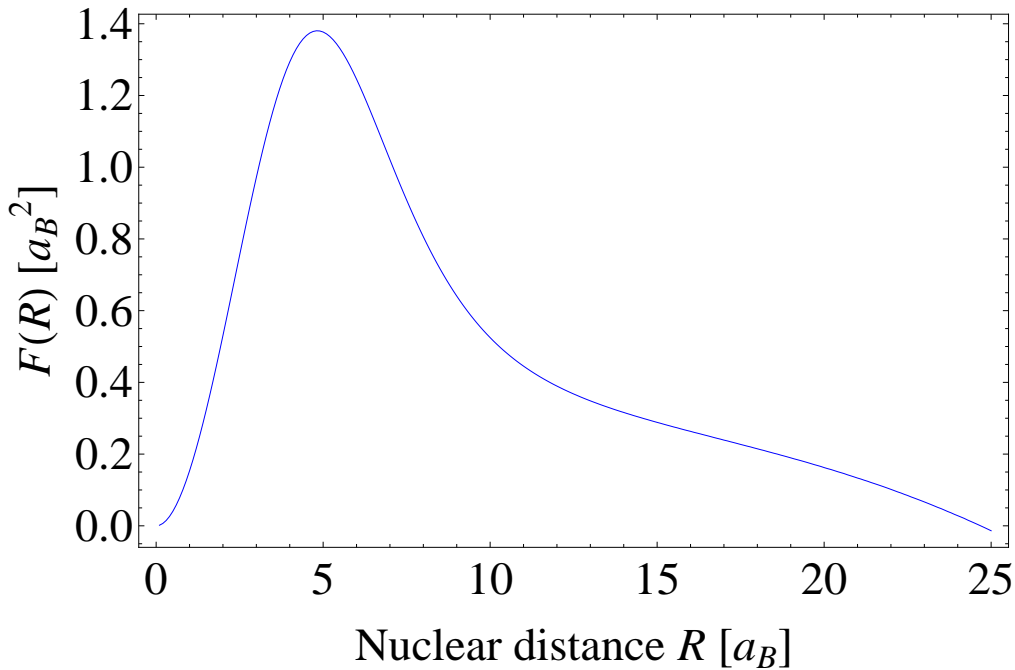}\caption{\label{fig:function
N} Plot of the function $F(R)$ (in atomic units). }
\end{figure}

\subsection{Comparison with previous work}

The values of $M(R)$ for homonuclear ions, calculated with our numerical
values of the function $F(R)$, agree with the results of Refs.~\cite{Bates and Poots}
(Table 1 therein), \cite{Bishop and Lam 1988} (Table 2 therein) and
\cite{Montgomery}. Also, the values of the function $F(R)$ are essentially
identical to those extracted from Ref.~\cite{Peek} in their Table
II.

For $\Htwop$, our value $\overline{M}_{00,00}=1.63775$~at.u. is in
agreement with Ref.~\cite{Bishop and Lam 1987} (Table 1 therein)
and the numerically less accurate, older value of Ref.~\cite{Bates
and Poots} (Table 3 therein). Our values
$\overline{M}_{0L,0L}$,$L=0,\ldots10$, agree within 0.001 atomic
units with the more accurate values of Ref.~\cite{Bishop and Lam
1988} (Table 3 therein) computed with the adiabatic potential.

Concerning $\HDp$ the only previous calculation known to us is
Ref.~\cite{Bathia} (Table I). There, the definition of the
quadrupole moment is
$\langle\frac{1}{2}(R^{2}+r^{2}-3z^{2})\rangle$, the same as for
the homonuclear ions. Thus, the expression
$M(R)|_{BD}=R^{2}m_{1}m_{2}/m_{12}^{2}+F(R)$ was used, which
involves a different dependence on the nuclear masses as compared
with our Eq.~(\ref{eq:thitaBred}). Our definition of the
quadrupole moment for $\HDp$ is
$\langle\left(\Theta_{B}\right)_{zz}\rangle=\langle\left(\tilde{\Theta}_{B}\right)_{0}\rangle=$$\langle\frac{1}{2}(2a_{0}R^{2}+r^{2}-3z^{2})\rangle$.
Accordingly, our value of $\overline{M}_{00,00}=1.7409$~at.u.
differs from $\overline{M}_{00,00}|_{BD}=1.505729$~at.u. (table I
in Ref.~\cite{Bathia}). However, if we compute $M(R)|_{BD}$ with
the functions $\chi_{1s\sigma}^{00}(R)$ calculated in the present
work, we obtain a similar result,
$\overline{M}_{00,00}'=1.5042$~at.u., which clearly indicates that
the discrepancy is due to the different analytical expressions
used, not to different wave functions.

\subsection{The quadrupole coupling coefficients in the effective Hamiltonian
$E_{14}$}

Eqs.~(\ref{eq:E14g}, \ref{eq:thitaBred}) lead to the following
expression of the quadrupole coupling coefficients $E_{14}$ of the
effective Hamiltonian for the ro-vibrational state $(v,L)$ in the
adopted approximation:

\begin{equation}
E_{14}=e\,\frac{\sqrt{6}}{3(2L-1)(2L+3)}\,\overline{M}_{vL,vL}.\label{eq:E14-BOfinal}
\end{equation}

Tables \ref{tab:Numerical-values-of E14},
\ref{tab:Numerical-values-of E14h2} and
\ref{tab:Numerical-values-of E14-1} list the values of $E_{14}$
for 99 ro-vibrational states of $\HDp$, $\Htwop$ and $\Dtwop$,
respectively, calculated using Eq.~(\ref{eq:E14-BOfinal}). Note
the slow increase of $E_{14}$ with $v$ and the stronger decrease
with $L$. The entries with $L=0$ are not of relevance in the
following, but are given for completeness since they are
proportional to the normalized quadrupole moment of the $L=0$
states, $\overline{M}_{v0,v0}=-9\, E_{14}(v,\,0)/\sqrt{6}\,.$

\begin{sidewaystable}
$
\begin{array}{r|@{\hspace{2mm}}r@{\hspace{2mm}}r@{\hspace{2mm}}r
@{\hspace{2mm}}r@{\hspace{2mm}}r@{\hspace{2mm}}r@{\hspace{2mm}}r
@{\hspace{2mm}}r@{\hspace{2mm}}r} L & \multicolumn{1}{c}{v=0} &
\multicolumn{1}{c}{1}  & \multicolumn{1}{c}{2} &
\multicolumn{1}{c}{3}  & \multicolumn{1}{c}{4} &
\multicolumn{1}{c}{5}  & \multicolumn{1}{c}{6} &
\multicolumn{1}{c}{7}  & \multicolumn{1}{c}{8} \\ \hline 0 &
-.3208($-$3) & -.3607($-$3) & -.4032($-$3) & -.4486($-$3) &
-.4972($-$3) & -.5494($-$3) & -.6056($-$3) & -.6664($-$3) &
-.7324($-$3) \\   1 & 0.1928($-$3) & 0.2168($-$3) & 0.2423($-$3) &
0.2696($-$3) & 0.2988($-$3) & 0.3302($-$3) & 0.3640($-$3) &
0.4005($-$3) & 0.4402($-$3) \\   2 & 0.4609($-$4) & 0.5180($-$4) &
0.5790($-$4) & 0.6441($-$4) & 0.7139($-$4) & 0.7888($-$4) &
0.8693($-$4) & 0.9565($-$4) & 0.1051($-$3) \\   3 & 0.2163($-$4) &
0.2430($-$4) & 0.2716($-$4) & 0.3021($-$4) & 0.3348($-$4) &
0.3699($-$4) & 0.4077($-$4) & 0.4485($-$4) & 0.4930($-$4) \\   4 &
0.1273($-$4) & 0.1430($-$4) & 0.1598($-$4) & 0.1777($-$4) &
0.1969($-$4) & 0.2176($-$4) & 0.2398($-$4) & 0.2638($-$4) &
0.2900($-$4) \\   5 & 0.8454($-$5) & 0.9495($-$5) & 0.1061($-$4) &
0.1179($-$4) & 0.1307($-$4) & 0.1443($-$4) & 0.1591($-$4) &
0.1750($-$4) & 0.1924($-$4) \\   6 & 0.6059($-$5) & 0.6803($-$5) &
0.7596($-$5) & 0.8446($-$5) & 0.9355($-$5) & 0.1033($-$4) &
0.1139($-$4) & 0.1253($-$4) & 0.1377($-$4) \\   7 & 0.4580($-$5) &
0.5140($-$5) & 0.5738($-$5) & 0.6377($-$5) & 0.7062($-$5) &
0.7799($-$5) & 0.8595($-$5) & 0.9456($-$5) & 0.1040($-$4) \\   8 &
0.3601($-$5) & 0.4040($-$5) & 0.4508($-$5) & 0.5009($-$5) &
0.5546($-$5) & 0.6124($-$5) & 0.6747($-$5) & 0.7424($-$5) &
0.8164($-$5) \\   9 & 0.2920($-$5) & 0.3273($-$5) & 0.3651($-$5) &
0.4056($-$5) & 0.4490($-$5) & 0.4957($-$5) & 0.5461($-$5) &
0.6010($-$5) & 0.6609($-$5) \\  10 & 0.2426($-$5) & 0.2718($-$5) &
0.3031($-$5) & 0.3365($-$5) & 0.3724($-$5) & 0.4111($-$5) &
0.4530($-$5) & 0.4985($-$5) & 0.5483($-$5)
\end{array}
$\caption{\label{tab:Numerical-values-of E14}Numerical values of the coefficients
$E_{14}$ of the effective Hamiltonian, Eq.~(\ref{eq:heff}), for
some ro-vibrational states $(v,\, L)$ of $\HDp$, with units MHz~m\textsuperscript{2}/GV.
The notation $a(-b)$ stands for $a\times10^{-b}$. In order to convert
the values to atomic units ($e\, a_{B}^{2})$, multiply by 1476.87.}
\end{sidewaystable}

\begin{sidewaystable}
$
\begin{array}{r|@{\hspace{2mm}}r@{\hspace{2mm}}r@{\hspace{2mm}}r
@{\hspace{2mm}}r@{\hspace{2mm}}r@{\hspace{2mm}}r@{\hspace{2mm}}r
@{\hspace{2mm}}r@{\hspace{2mm}}r}
 L & \multicolumn{1}{c}{v=0} & \multicolumn{1}{c}{1}  & \multicolumn{1}{c}{2} & \multicolumn{1}{c}{3}  & \multicolumn{1}{c}{4} & \multicolumn{1}{c}{5}  & \multicolumn{1}{c}{6} & \multicolumn{1}{c}{7}  & \multicolumn{1}{c}{8}
 \\
 \hline
 0 & $-$0.3018($-$3) & $-$0.3448($-$3) & $-$0.3910($-$3) & $-$0.4409($-$3) &
 $-$0.4948($-$3) & $-$0.5533($-$3) & $-$0.6172($-$3) & $-$0.6874($-$3) &
 $-$0.7652($-$3)   \\   1 &  0.1815($-$3) &  0.2074($-$3) &  0.2351($-$3) &
 0.2651($-$3) &  0.2975($-$3) &  0.3327($-$3) &  0.3711($-$3) &  0.4133($-$3) &
 0.4601($-$3)   \\   2 &  0.4343($-$4) &  0.4960($-$4) &  0.5624($-$4) &
 0.6340($-$4) &  0.7115($-$4) &  0.7956($-$4) &  0.8873($-$4) &  0.9883($-$4) &
 0.1100($-$3)   \\   3 &  0.2042($-$4) &  0.2331($-$4) &  0.2642($-$4) &
 0.2978($-$4) &  0.3342($-$4) &  0.3736($-$4) &  0.4167($-$4) &  0.4641($-$4) &
 0.5168($-$4)   \\   4 &  0.1205($-$4) &  0.1375($-$4) &  0.1558($-$4) &
 0.1756($-$4) &  0.1970($-$4) &  0.2202($-$4) &  0.2456($-$4) &  0.2736($-$4) &
 0.3046($-$4)   \\   5 &  0.8022($-$5) &  0.9151($-$5) &  0.1037($-$4) &
 0.1168($-$4) &  0.1310($-$4) &  0.1464($-$4) &  0.1633($-$4) &  0.1819($-$4) &
 0.2026($-$4)   \\   6 &  0.5769($-$5) &  0.6577($-$5) &  0.7448($-$5) &
 0.8388($-$5) &  0.9406($-$5) &  0.1051($-$4) &  0.1173($-$4) &  0.1306($-$4) &
 0.1455($-$4)   \\   7 &  0.4377($-$5) &  0.4987($-$5) &  0.5645($-$5) &
 0.6355($-$5) &  0.7124($-$5) &  0.7962($-$5) &  0.8881($-$5) &  0.9895($-$5) &
 0.1103($-$4)   \\   8 &  0.3456($-$5) &  0.3935($-$5) &  0.4452($-$5) &
 0.5010($-$5) &  0.5615($-$5) &  0.6275($-$5) &  0.6999($-$5) &  0.7800($-$5) &
 0.8698($-$5)   \\   9 &  0.2814($-$5) &  0.3202($-$5) &  0.3621($-$5) &
 0.4073($-$5) &  0.4564($-$5) &  0.5100($-$5) &  0.5689($-$5) &  0.6342($-$5) &
 0.7075($-$5)   \\  10 &  0.2349($-$5) &  0.2671($-$5) &  0.3019($-$5) &
 0.3394($-$5) &  0.3803($-$5) &  0.4249($-$5) &  0.4740($-$5) &
 0.5286($-$5)
 &
 0.5902($-$5)
\end{array}
$\caption{\label{tab:Numerical-values-of E14h2}Same as Table \ref{tab:Numerical-values-of E14},
but for $\Htwop$.}
\end{sidewaystable}

\begin{sidewaystable}
$
\begin{array}{r|@{\hspace{2mm}}r@{\hspace{2mm}}r@{\hspace{2mm}}r  @{\hspace{2mm}}r@{\hspace{2mm}}r@{\hspace{2mm}}r@{\hspace{2mm}}r  @{\hspace{2mm}}r@{\hspace{2mm}}r}
L & \multicolumn{1}{c}{v=0} & \multicolumn{1}{c}{1}  & \multicolumn{1}{c}{2} & \multicolumn{1}{c}{3}  & \multicolumn{1}{c}{4} & \multicolumn{1}{c}{5}  & \multicolumn{1}{c}{6} & \multicolumn{1}{c}{7}  & \multicolumn{1}{c}{8} \\ \hline
  0 & $-$0.2957($-$3) & $-$0.3254($-$3) & $-$0.3568($-$3) & $-$0.3898($-$3) & $-$0.4246($-$3) & $-$0.4613($-$3) & $-$0.5002($-$3) & $-$0.5414($-$3) & $-$0.5852($-$3) \\   1 &  0.1776($-$3) &  0.1955($-$3) &  0.2143($-$3) &  0.2341($-$3) &  0.2550($-$3) &  0.2771($-$3) &  0.3004($-$3) &  0.3252($-$3) &  0.3515($-$3) \\   2 &  0.4240($-$4) &  0.4666($-$4) &  0.5114($-$4) &  0.5587($-$4) &  0.6085($-$4) &  0.6612($-$4) &  0.7169($-$4) &  0.7759($-$4) &  0.8386($-$4) \\   3 &  0.1986($-$4) &  0.2185($-$4) &  0.2395($-$4) &  0.2616($-$4) &  0.2849($-$4) &  0.3096($-$4) &  0.3356($-$4) &  0.3632($-$4) &  0.3926($-$4) \\   4 &  0.1166($-$4) &  0.1283($-$4) &  0.1406($-$4) &  0.1536($-$4) &  0.1673($-$4) &  0.1817($-$4) &  0.1970($-$4) &  0.2132($-$4) &  0.2304($-$4) \\   5 &  0.7722($-$5) &  0.8494($-$5) &  0.9307($-$5) &  0.1016($-$4) &  0.1107($-$4) &  0.1202($-$4) &  0.1303($-$4) &  0.1410($-$4) &  0.1524($-$4) \\   6 &  0.5515($-$5) &  0.6065($-$5) &  0.6645($-$5) &  0.7255($-$5) &  0.7900($-$5) &  0.8580($-$5) &  0.9301($-$5) &  0.1006($-$4) &  0.1088($-$4) \\   7 &  0.4152($-$5) &  0.4565($-$5) &  0.5000($-$5) &  0.5459($-$5) &  0.5943($-$5) &  0.6454($-$5) &  0.6995($-$5) &  0.7570($-$5) &  0.8181($-$5) \\   8 &  0.3250($-$5) &  0.3573($-$5) &  0.3912($-$5) &  0.4270($-$5) &  0.4648($-$5) &  0.5047($-$5) &  0.5470($-$5) &  0.5919($-$5) &  0.6397($-$5) \\   9 &  0.2622($-$5) &  0.2881($-$5) &  0.3155($-$5) &  0.3442($-$5) &  0.3746($-$5) &  0.4068($-$5) &  0.4408($-$5) &  0.4769($-$5) &  0.5154($-$5) \\  10 &  0.2167($-$5) &  0.2381($-$5) &  0.2605($-$5) &  0.2843($-$5) &  0.3093($-$5) &  0.3358($-$5) &  0.3638($-$5) &  0.3936($-$5) &  0.4254($-$5)
\end{array}
$\caption{\label{tab:Numerical-values-of E14-1}Same as Table \ref{tab:Numerical-values-of E14},,
but for $\Dtwop$ .}
\end{sidewaystable}

\section{The quadrupole shift in $\HDp$}

\subsection{Generalities}

We denote by $E^{vLnJ_{z}}(\mathbf{B},Q)$ the energy of the
hyperfine state $|vLnJ_{z}(\mathbf{B},Q)\rangle$ of $\HDp$ in a
magnetic field $\mathbf{B}$ and in an electric field gradient $Q$.
Because of the spin interactions these states are not in general
eigenstates of the operators ${\bf F}^{2},\,{\bf S}^{2}$ and ${\bf
J}^{2}$ and the quantum numbers $F,\, S$ and $J$ associated with
them are not exact quantum numbers, but for weak fields may be
considered as approximate quantum numbers. For given values of $L$
and $J_{z}$, the number $N(L,\, J_{z})$
of\lyxdeleted{schiller}{Sat Oct 26 22:43:05 2013}{ } eigenstates
$|vLnJ_{z}(\mathbf{B},Q)\rangle$ is equal to the number of the
combinations of quantum numbers $(F,\, S,\, J)$ in the spin
coupling scheme of Ref.\cite{Zee2} allowed by angular momentum
algebra. We therefore use the index $n=1,\,2,...,\, N(L,\,
J_{z})$, which enumerates the possible combinations of $(F,\, S,\,
J)$, to label the spin content of $|vLnJ_{z}(\mathbf{B},Q)\rangle$
and associate each value of $n$ with the set of values of the
approximate quantum numbers:
$n\Leftrightarrow(F_{n},S_{n},J_{n})$. Note that $J$ is exact in
absence of external fields, and $J_{z}$ is exact if the axial
symmetry is conserved.

We consider in the following three levels of perturbation calculations,
which are all restricted to a given level $(v,\, L)$: (1) diagonalizing
the whole effective Hamiltonian $H_{{\rm eff}}^{{\rm tot+Q}}$ between
angular momentum basis states; (2) diagonalizing the matrix of the
quadrupole interaction $V^{\mathrm{Q}}$ between eigenstates of the
effective Hamiltonian $H_{{\rm eff}}^{{\rm tot}}$ that includes magnetic,
but not the quadrupole interaction; (3) compute the expectation value
of the quadrupole interaction. We then show that the latter approximation
is sufficient.

\subsection{Diagonalization of the effective hamiltionian in a state $(v,\, L)$ \label{sub:Diagonalization-of-the effective hamiltonian}}

The energies $E^{vLnJ_{z}}(\mathbf{B},Q)$ are defined as
eigenvalues of the matrix of the effective spin Hamiltonian
$H_{{\rm eff}}^{{\rm tot+Q}}$ of Eq.~(\ref{eq:heff}) in the
subspace of states with fixed values of $v$ and $L$. This matrix
has dimension $(2S_{p}+1)(2S_{d}+1)(2S_{e}+1)(2L+1)$ squared
($S_p$, $S_d$, $S_e$ being the spins of the three particles), i.e.
$12(2L+1)\times12(2L+1)$. The matrix elements of the spin
interaction operators (the first 9 terms of $H_{{\rm eff}}^{{\rm
tot+Q}}$) were computed in \cite{prl06} I changed ``give'' to
``computed'' {*}{*}{*}{*} actually, we did not give the matrix
elements, just the hamiltonian {*}{*}{*}, and those of the
interactions with external magnetic field $V^{\mathrm{mag}}$ (next
4 terms) in \cite{Zee2}. With account of Eqs.~(\ref{eq:thitaBred})
and (\ref{eq:E14-BOfinal}), the matrix elements of the projection
$V^{\mathrm{Q}}$ of $\Delta H_{Q}$ on the subspace of a state with
fixed values of $v$ and $L$ have the form:

\begin{eqnarray}
\left\langle vLF'S'J'J'_{z}|V^{Q}|vLFSJJ_{z}\right\rangle  & = & E_{14}\,\delta_{S'S}\delta_{F'F}\,(-1)^{J'+S+L}\langle L||(\mathbf{L}\otimes\mathbf{L})^{(2)}||L\rangle\times\label{eq:delta_VQ0}\\
 & \times & \sqrt{2J+1}\,\left\{ \begin{array}{ccc}
L & 2 & L\\
J' & S & J
\end{array}\right\} \sum_{q}\,\tilde{Q}^{q}(\boldsymbol{\mathrm{R}}_{B})\, C_{JJ_{z},2q}^{J'J_{z}'}\ .\nonumber
\end{eqnarray}

Note that they vanish for $L=0$ levels.

\subsection{Diagonalization of the electric quadrupole hamiltionian in the space
of Zeeman hyperfine states}

Comparison of the values of $E_{14}$ with the values of the
coefficients $E_{k},\, k=10,\ldots,13$ of the effective
Hamiltonian for the Zeeman effect \cite{Zee2} shows that, for the
electric field gradients and magnetic fields of interest here, the
quadrupole shift $\Delta E_{{\rm
Q}}^{vLnJ_{z}}=E^{vLnJ_{z}}(\mathbf{B},Q)-E^{vLnJ_{z}}(\mathbf{B},0)$
is for the majority of levels much smaller than the Zeeman shift
$E^{vLnJ_{z}}(\mathbf{B},0)-E^{vLnJ_{z}}(\mathbf{\mathrm{0}},0)$.
Even the hyperfine states least sensitive to magnetic fields,
those with $J_{z}=0$ (having only a quadratic Zeeman shift),
exhibit at 1~G a typical shift of a few kHz or more, occasionally
only tens of Hz, while the electric quadrupole shift, in a
10\textsuperscript{8} V/m\textsuperscript{2} gradient, is of the
order of 100 Hz. Therefore, for sufficiently large magnetic fields
the electric quadrupole shift can conveniently be evaluated as a
perturbation to the Zeeman-shifted hyperfine energy levels by
diagonalizing the matrix of $V^{\mathrm{Q}}$, Eq.~(\ref{eq:heff}),
in the basis of the Zeeman-shifted hyperfine states
$|vLnJ_{z}(\mathbf{B},0)\rangle$, calculated as eigenvectors of
the spin and magnetic interaction part $H_{{\rm eff}}^{{\rm
tot+Q}}=H_{{\rm eff}}^{{\rm hfs}}+V^{{\rm mag}}$ of the effective
Hamiltonian of Eq.~(\ref{eq:heff}):
\begin{eqnarray}
\left\langle vLn'J_{z}'(\mathbf{B},0)|V^{\mathrm{Q}}|vLnJ_{z}(\mathbf{B},0)\right\rangle  & = & E_{14}\,\langle L||(\mathbf{L}\otimes\mathbf{L})^{(2)}||L\rangle\times\label{eq:vQmatrix}\\
 & \times & \sum_{q}\tilde{Q}^{q}(\boldsymbol{\mathrm{R}}_{B})\sum_{FSJ'J}(-1)^{S+J'+L}\sqrt{2J+1}\times\nonumber \\
 & \times & \left\{ \begin{array}{ccc}
L & 2 & L\\
J' & S & J
\end{array}\right\} C_{JJ_{z},2q}^{J'J'_{z}}\,\beta_{FSJ'}^{vLn'J'_{z}}(\mathbf{B})\beta_{FSJ}^{vLnJ_{z}}(\mathbf{B})\,,\nonumber
\end{eqnarray}
where $\beta_{FSJ}^{vLnJ_{z}}(\mathbf{B})$ are the expansion coefficients
of the hyperfine states in presence of a magnetic field $\mathbf{B}$
in the field-free basis set $\left\{ |vLFSJJ_{z}\rangle\right\} $
\cite{Zee2}:

\begin{equation}
|vLnJ_{z}(\mathbf{B},0)\rangle=\sum_{F'S'J'}\beta_{F'S'J'}^{vLnJ_{z}}(\mathbf{B})|vLF'S'J'J_{z}\rangle\,.\label{eq:zee2exp}
\end{equation}
Note that in Eq.~(\ref{eq:zee2exp}) there is no summation over the
angular momentum projection $J_{z}$, since it remains a good quantum
number in a homogeneous magnetic field. The computational advantage
of evaluating the electric quadrupole shift $\Delta E_{{\rm Q}}^{vLnJ_{z}}$
by diagonalizing the matrix of $V^{\mathrm{Q}}$ in Eq.~(\ref{eq:vQmatrix})
instead of $H_{{\rm eff}}^{{\rm tot+Q}}$ is that no precision is
lost in the subtraction $E^{vLnJ_{z}}(\mathbf{B},Q)-E^{vLnJ_{z}}(\mathbf{B},0)$.
Note again, that the matrix element in Eq.~(\ref{eq:vQmatrix}) vanishes
for $L=0$.

\subsection{First-order perturbation theory}

In first order of perturbation theory the quadrupole shift is given
by the diagonal matrix element of $V^{\mathrm{Q}}$,

\begin{eqnarray}
\Delta E_{{\rm Q,diag}}^{vLnJ_{z}} & = & \left\langle vLnJ_{z}(\mathbf{B},0)|V^{\mathrm{Q}}|vLnJ_{z}(\mathbf{B},0)\right\rangle \nonumber \\
 & = & E_{14}\tilde{Q}^{0}(\boldsymbol{\mathrm{R}}_{B})\left\langle vLnJ_{z}(\mathbf{B},0)|(\mathbf{L}\otimes\mathbf{L})_{0}^{(2)}|vLnJ_{z}(\mathbf{B},0)\right\rangle \nonumber \\
 & = & E_{14}\,\langle L||(\mathbf{L}\otimes\mathbf{L})^{(2)}||L\rangle\:\tilde{Q}^{0}(\boldsymbol{\mathrm{R}}_{B})\times\label{eq:vQdiag1}\\
 & \times & \sum_{FSJ'J}(-1)^{S+J'+L}\,\sqrt{2J+1}\, C_{JJ_{z},20}^{J'J_{z}}\,\beta_{FSJ'}^{vLnJ_{z}}(\mathbf{B})\beta_{FSJ}^{vLnJ_{z}}(\mathbf{B})\left\{ \begin{array}{ccc}
L & 2 & L\\
J' & S & J
\end{array}\right\} \nonumber \\
\nonumber \\
\nonumber
\end{eqnarray}
to which only the longitudinal component $Q_{zz}=\tilde{Q}^{0}({\bf R}_{B})$
of the electric field gradient contributes; since $Q_{zz}$ does not
mix states with different values of $J_{z}$, $J_{z}$ remains a good
quantum number in this case. The transversal components that couple
states with differents values of $J_{z}$ contribute in second order
of perturbation theory only; for the electric and magnetic fields
of interest the second-order effects are below 0.1 Hz and will be
neclected in what follows.

We may rewrite the above equation in a simplified notation,

\begin{equation}
\Delta E_{{\rm Q,diag}}^{vLFSJJ_{z}}=\sqrt{\frac{3}{2}}\, E_{14}(v,\, L)\, Q_{zz}\left\langle vLFSJ\, J_{z}|\, L_{z}^{2}-\frac{1}{3}{\bf L}^{2}|vLFSJ\, J_{z}\right\rangle \,.
\end{equation}

From the property of the 6-j symbols we see that the shift vanishes
if $L=0$. Furthermore, in the limit of zero magnetic field $B$,
\begin{equation}
\beta_{FSJ}^{vLnJ_{z}}(\mathbf{B})\simeq\beta_{FSJ}^{vLnJ_{z}}(0)=\delta_{JJ_{n}}\beta_{FSJ}^{vL(F_{n}S_{n}J_{n})J_{z}}(0)\,.\label{eq:betaB}
\end{equation}

The sum over $J,\, J'$ in Eq.~(\ref{eq:vQdiag1}) is then proportional
to

\[
\left\{ \begin{array}{ccc}
L & L & 2\\
J & J & S'
\end{array}\right\} \,.
\]
This 6-j symbol vanishes for $J=J_{n}=0$ hyperfine levels. Therefore,
the shift nearly vanishes for such levels in the limit of small magnetic
field. Thus, for example, a transition $(v,L=0,n,\, J_{z})\rightarrow(v',L',n',\, J_{z}'=0)$
such that \lyxadded{mitko}{Fri Oct 04 19:40:25 2013}{$J_{n'}=0$}
is nearly free of quadrupole shift if the magnetic field is small.
Table~\ref{tab:rotational transitions} below contains such a transition.

\subsection{Numerical example}

We have performed numerical diagonalization of the effective
hamiltonian Eq.~(\ref{eq:heff}), including hyperfine coupling,
Zeeman interaction and quadrupole interaction as described in
Sec.~\ref{sub:Diagonalization-of-the effective hamiltonian}. For
example, for the level $(v=0,\, L=1)$ in $B=1$~G, and a purely
longitudinal gradient
$Q=Q_{zz}=$~10\textsuperscript{8}~V/m\textsuperscript{2}
 for a gradient with non-zero transversal
components $Q$ = $(\tilde{Q}^{-1}=100\times10^{8},\,
\tilde{Q}^{0}=10^{8})$~V/m\textsuperscript{2},
and for $\,(\tilde{Q}^{-2}=100\times10^{8},\,
\tilde{Q}^{0}=10^{8})$~V/m\textsuperscript{2} the largest relative
difference between the ``exact'' quadrupole shift $\Delta E_{{\rm
Q}}^{vLnJ_{z}}$ and the diagonal approximation $\Delta E_{{\rm
Q,diag}}^{vLnJ_{z}}$, Eq.~(\ref{eq:vQdiag1}), is $9\times10^{-4}$.
The maximum absolute deviation is $9\times10^{-5}\,$Hz. For
concreteness, we have also studied the effect of reducing the
magnetic field  from 1 G to 0.1 G for the $(0,3)$, $(3,4)$ and
$(5,4)$ levels that support some of the metrologically interesting
transitions listed in Table 7. All Zeeman states exhibit very
small relative and very small absolute differences ($<0.06\,$Hz)
between the full diagonalization value
 (when ${\tilde Q}_{\pm2}={\tilde Q}_{\pm1}=\tilde{Q}_0$ was set) and the
expectation value results (which takes only $\tilde{Q}_0$ into
account) also in 0.1 G, except for the $J_z \ne0$ hyperfine Zeeman
states of those two particular hyperfine levels that also contain
the particularly favorable $J_z=0\rightarrow J'_z=0$ transition
with -2.3 Hz Zeeman shift (Table 7). For the $J'_z\ne0$ states the
absolute difference increases from a maximum of 0.4 Hz to a
maximum of 2.5 Hz when $B$ is reduced to $0.1\,$G. For the
$J'_z=0$ states it does not exceed 0.25\,Hz even in 0.1 G. These
differences are related to the small Zeeman splitting in these
particular hyperfine levels. However, these differences do not
affect the discussion and conclusions below. There is no
difference if ${\tilde Q}_0$ is the only nonzero component.
Thus the diagonal approximation produces - within the adopted
accuracy - essentially the same numerical values of the quadrupole
shift as the full diagonalization of the effective hamiltonian.

\subsection{The shift of the stretched states}

A special case is the stretched states. For any rovibrational level
$(v,\, L)$, these are the two states with maximum total angular momentum
and projection,

\[
|vLn_{s}J_{z}({\bf B},0)\rangle=|v,\, L,\, F=1,\, S=2,\, J=L+2,\, J_{z}=\pm(L+2)({\bf B},0)\rangle\,,
\]
introduced in Ref.~\cite{Zee2}. The expansion Eq.~(\ref{eq:zee2exp})
of these for any magnetic field strength contains only a single nonzero
coefficient, $\beta_{FSJ}^{vLn_{s}J_{z}=\pm J}(\mathbf{B})=\delta_{F1}\delta_{S2}\delta_{JL+2}$.Using
this, we obtain the simple expression for both stretched states:

\begin{eqnarray}
\Delta E_{{\rm Q,diag}}^{vLn_{s}J_{z}=\pm(L+2)}(B) & =\frac{L\,(2\, L-1)}{\sqrt{6}}\, E_{14}\, Q_{zz}\,.
\end{eqnarray}
The shift is equal for both stretched states and independent of magnetic
field strength.

\section{Numerical results for $\HDp$}

\subsection{Energy shifts}

To illustrate the magnitude of the electric quadrupole shift, we list
in Table~\ref{tab:Shifts of energy levels} the quadrupole shifts
$\Delta E_{{\rm Q,diag}}^{vLnJ_{z}}(\mathbf{B})$ of the hyperfine
energy levels of the initial and final states, $(v\!=\!0,\, L\!=\!1)$
and $(v\!=\!4,\, L\!=\!2)$, of a particular one-photon rovibrational
transition in $\HDp,$ discussed in detail in Ref.~\cite{Zee2}.
We choose a value $Q_{zz}=0.1$~GV~m$^{-2}$ which could be present
in a linear ion trap in which one $\HDp$ ion and one Be\textsuperscript{+}ion
(for sympathetic cooling and quantum logic interrogation) are located
at a few \textmu{}m distance. Comparison with Table 2 of Ref.~\cite{Zee2}
shows that the quadrupole shift is typically orders of magnitude smaller
than the Zeeman shift. We emphasize that the quadrupole shift of a
given hyperfine state does depend on the magnetic field strength,
although the dependence is weak for the majority of the states (at
the field value assumed in the Table). It is useful to compare the
values with those relevant for a particular atomic ion used for atomic
ion clocks: the upper level of the octupole transition of \textsuperscript{171}Yb\textsuperscript{+}
has an electric quadrupole shift of 2~Hz in the same gradient, at
a transition frequency of 642~THz\lyxadded{schiller}{Sat Oct 26 17:37:40 2013}{
\cite{Huntemann-1}}.

\begin{table}
$\begin{array}{|c|rrrrrrrrr|}
 \hline \text{state} &  &  &  &  & J_{z} &  &  &  & \\[-.3cm]
 (FSJ) & -4 & -3 & -2 & -1 & 0 & 1 & 2 & 3 & 4\\
 \hline\hline \text{\ensuremath{(124)}} & {\bf 17.5} & 5.8 & -3.3 & -9.5 & -12.4 & -11.6 & -6.7 & 2.9 & {\bf 17.5}\\[-.3cm]
 & \text{} & 17.5 & 0.4 & -10.1 & -14.0 & -10.8 & -0.4 & 17.5 & \text{}\\[-.3cm]
 \text{\ensuremath{(113)}} & \text{} & 17.5 & 0.2 & -10.3 & -14.0 & -10.6 & -0.2 & 17.5 & \text{}\\[-.3cm]
 \text{\ensuremath{(123)}} & \text{} & 2.9 & -1.2 & -3.3 & -3.6 & -2.1 & 1.1 & 5.9 & \text{}\\[-.3cm]
 & \text{} & \text{} & 8.3 & -4.0 & -8.7 & -4.7 & 9.2 & \text{} & \text{}\\[-.3cm]
 \text{(102)} & \text{} & \text{} & 16.0 & -8.0 & -16.1 & -8.0 & 16.1 & \text{} & \text{}\\[-.3cm]
 \text{\ensuremath{(112)}} & \text{} & \text{} & 9.6 & -3.8 & -9.7 & -5.8 & 9.9 & \text{} & \text{}\\[-.3cm]
 \text{\ensuremath{(122)}} & \text{} & \text{} & -3.8 & 1.7 & 3.3 & 1.6 & -2.8 & \text{} & \text{}\\[-.3cm]
 \text{\ensuremath{(011)}} & \text{} & \text{} & \text{} & 5.4 & -12.2 & 6.8 & \text{} & \text{} & \text{}\\[-.3cm]
 \text{\ensuremath{(111)}} & \text{} & \text{} & \text{} & 4.2 & -10.9 & 6.5 & \text{} & \text{} & \text{}\\[-.3cm]
 \text{\ensuremath{(121)}} & \text{} & \text{} & \text{} & -5.9 & 10.9 & -5.0 & \text{} & \text{} & \text{}\\[-.3cm]
 \text{\ensuremath{(120)}} & \text{} & \text{} & \text{} & \text{} & \underline{0.05} & \text{} & \text{} & \text{} & \text{}
 \\\hline \end{array}
 $\vskip 10pt$
 \begin{array}{|c|rrrrrrr|}
 \hline \text{state} &  &  &  & J_{z} &  &  & \\[-.3cm]
 \text{\ensuremath{(FSJ)}} & -3 & -2 & -1 & 0 & 1 & 2 & 3\\
 \hline\hline \text{(123)} & {\bf 7.9} & -4.9 & -7.5 & -5.6 & -1.5 & 3.2 & {\bf 7.9}\\[-.3cm]
 \text{(012)} & \text{} & 7.9 & -3.4 & -7.8 & -4.5 & 7.9 & \text{}\\[-.3cm]
 \text{(112)} & \text{} & 7.9 & -3.8 & -7.9 & -4.1 & 7.9 & \text{}\\[-.3cm]
 \text{(122)} & \text{} & -3.0 & 6.6 & 7.2 & 0.8 & -11.1 & \text{}\\[-.3cm]
 (011) &  &  & -4.5 & 7.8 & -3.4 &  & \\[-.3cm]
 \text{(101)} & \text{} & \text{} & 6.6 & -13.1 & 6.6 & \text{} & \text{}\\[-.3cm]
 \text{(111)} & \text{} & \text{} & -2.5 & 4.6 & -2.2 & \text{} & \text{}\\[-.3cm]
 \text{(121)} & \text{} & \text{} & 0.6 & -0.9 & 0.4 & \text{} & \text{}\\[-.3cm]
 \text{(010)} & \text{} & \text{} & \text{} & \underline{0.02} & \text{} & \text{} & \text{}\\[-.3cm]
 \text{(110)} & \text{} & \text{} & \text{} & \underline{0.02} & \text{} & \text{} & \text{}
 \\
 \hline
 \end{array}
 $\caption{\label{tab:Shifts of energy levels}
Quadrupole shifts $\Delta E_{{\rm Q}}^{vLnJ_{z}}(B)$ (in Hz) of
the hyperfine states $n=(FSJ)$ in the $(v,\, L)=(4,\,2)$ (top) and
$(0,\,1)$ (bottom) rovibrational states of $\HDp$. The magnetic
field is $B=1$~G and the electric field gradient
$Q=\tilde{Q}^{0}=Q_{zz}=0.1$~GV m$^{-2}$, with the only
nonvanishing component along the magnetic field
\lyxdeleted{schiller}{Sun Oct 06 17:28:33 2013}{}
\lyxdeleted{mitko}{Fri Oct 04 19:43:36 2013}{to $\mathbf{B}$}. The
underlined numbers correspond to hyperfine states with $J=0$, for
which the quadrupole shift vanishes in the limit of vanishing
magnetic field. The bold numbers correspond to stretched states.}
\end{table}

\subsection{Metrologically interesting transitions}

Since the Zeeman shift is the dominant shift, we have searched for
transitions with small Zeeman shifts of the transition frequencies
when the magnetic field is moderate (1~G), and report below their
electric quadrupole shifts $\Delta f_{Q}=(\Delta E_{{\rm
Q,diag}}^{v'L'n'J{}_{z}'}(\mathbf{B})-\Delta E_{{\rm
Q,diag}}^{vLnJ_{z}}(\mathbf{B}))/h$. For simplicity, we confined
the search to the range $v'\le5$\lyxadded{schiller}{Sat Oct 26
13:42:13 2013}{ \cite{Korobov and Bekbaev private communication
2013}}, implying transition wavelengths larger than approximately
1.1 $\mu$m.The search also found transitions with a small Zeeman
shift at 1~G which is of spurious origin, the transition not
actually being weakly dependent on the magnetic field. Such
transitions are not discussed further. This leaves essentially two
types of transitions (exceptions are mentioned below):

(i) of the type $J_{z}=0\rightarrow J'_{z}=0$, characterized by a
quadratic Zeeman shift, and

(ii) transitions between stretched states. For any pair of rovibrational
levels $(v,\, L)$, $(v',\, L')$, these are the two transitions

\begin{eqnarray*}
(v,\, L,\, F & = & 1,\, S=2,\, J=L+2,\, J_{z}=\,\,\, J)\rightarrow(v',\, L',\, F'=1,\, S'=2,\, J'=L'+2,\, J'_{z}=\,\,\,\, J'),\\
(v,\, L,\, F & = & 1,\, S=2,\, J=L+2,\, J_{z}=-J)\rightarrow(v',\, L',\, F'=1,\, S'=2,\, J'=L'+2,\, J'_{z}=-J')\,.
\end{eqnarray*}
Their favorable metrological properties have been discussed in Ref.~\cite{Zee2}.
Basically, since the Zeeman shift of the transition doublet is strictly
linear, one has the possibility of nulling the effect of the magnetic
field by measuring both transition frequencies (at any actual value
of the magnetic field) and then computing the average value. However,
the electric quadrupole shift is equal for both transitions in the
doublet, so no simple cancellation occurs.

\subsection{Radio-frequency transitions\label{sub:Radio-frequency-transitions}}

Magnetic (M1) hyperfine transitions within rovibrational levels having
rotational angular momentum $L=0$ are free of electric quadrupole
shifts. Unfortunately, all M1 transitions in the rovibrational ground
state $(v=0,\, L=0)$, which is well accessible experimentally, have
comparatively large Zeeman shifts.

It may be of interest to measure hyperfine transitions in levels with
nonzero $L$, in order to test $L$-dependent contributions to their
frequencies. For this purpose, Table \ref{table:RF transitions shifts}
shows a list of transitions between hyperfine states selected with
the criterium of less than 0.1~kHz Zeeman shift at 1~G for individual
transitions with quadratic Zeeman effect and less than 0.6~kHz shift
of the mean frequency of transition pairs. We have included transitions
with both small and large RF frequency. No selection was performed
with respect to the electric quadrupole shift because the criterium
of small Zeeman shifts is regarded as more important for experimental
reasons. In the search, we confined ourselves to the range $v=0,\,1$,
and $L=0,\,1,\,2$ in order to limit the number of results.

We find a substantial number of $J_{z}=0\rightarrow J'_{z}=0$ transitions
with Zeeman shifts of approximately 0.2 to 0.5~kHz at 1~G. A particularly
low Zeeman shift (3~Hz in 1~G, 0.3~Hz in 0.5 G) occurs for the
947.6~MHz hyperfine transition in $(v=1,\, L=1).$ This shift is
closely quadratic in $B$ only for $B<0.4$~G. Since the electric
quadrupole shift is also low, $-1.1$~Hz, the transition is an interesting
candidate for a precision test of the hyperfine hamiltonian. Note,
however, that this rovibrational level is an excited one, with finite
spontaneous lifetime (55~ms), giving rise to a natural broadening
of the transition of 6~Hz. Suppose that we can measure an RF transition
frequency with a resolution equal to 1~\% of the natural linewidth,
i.e. 0.06~Hz. By measuring the transition frequency as a function
of the magnetic field, it should be feasible to reduce the Zeeman
effect uncertainty to below 0.03~Hz.

Furthermore, a number of transition pairs exist (including in $v=0$)
which have large but nearly opposite Zeeman shifts, with a modest
mean shift. Examples with particularly low mean shift, from 2 to 80
Hz at 1~G, are shown in the table. We note that due to the nearly
complete cancellation of the opposite shifts, the mean shifts should
be considered as indicative only. It should be noted that small magnetic
field gradients in the ion trap will cause inhomogeneous broadening
of these RF transitions if spectroscopy is performed on ensembles
of ions.

\begin{sidewaystable}
\[
 \begin{array}{|c|cccr|cccr|r|r|r|r|r|r|r|r|}
 \hline (\text{\textit{\ensuremath{v}}},\text{\textit{\ensuremath{L}}})
 & F' & S' & J' & J_{z}' & F & S & J & J_{z} & f_{0}\text{(1 G)} & \text{rel.}
 & \Delta\text{\textit{\ensuremath{f}}}_{B} & \Delta\text{\textit{\ensuremath{f}}}_{Q}\text{(1 G)}
 & \text{(\ensuremath{\Delta}}\text{\textit{\ensuremath{E}}}_{Q})_{u}
 & \text{(\ensuremath{\Delta}}\text{\textit{\ensuremath{E}}}_{Q})_{l}
 & \Delta\text{\textit{\ensuremath{\alpha^{(t)}}}} & \Delta\text{\textit{\ensuremath{\alpha}}}^{(l)}\\
 \text{} & \text{} & \text{} & \text{} & \text{} & \text{} & \text{} & \text{} & \text{} & \text{[MHz]} & \text{int.}
 & \text{[Hz]} & \text{[Hz]} & \text{[Hz]} & \text{[Hz]} & \text{[at.u.]} & \text{[at.u.]}\\
 \hline\hline \text{(0, 1)} & 1 & 2 & 1 & 0 & 0 & 1 & 0 & 0 & 969. & 0.787 & 422 & -1.0 & -0.9 & 0.0 & 7.1 & -14.1\\[-.3cm]
 \text{(0, 2)} & 1 & 2 & 1 & -1 & 1 & 0 & 2 & -2 & 184.1 & 0.004 & -1340044 & -13.7 & -3.6 & 10.1 & 33.1 & -66.1\\[-.3cm]
 \text{(0, 2)} & 1 & 2 & 1 & 1 & 1 & 0 & 2 & 2 & 186.8 & 0.003 & 1339960 & -13.3 & -3.1 & 10.1 & 33.1 & -66.1\\[-.3cm]
 \text{(0, 2)} & 1 & 2 & 2 & 0 & 1 & 1 & 3 & 0 & 97.3 & 0.029 & -150 & 11.1 & 2.1 & -9.0 & -27.2 & 54.3\\[-.3cm]
 \text{(0, 3)} & 1 & 1 & 4 & -4 & 0 & 1 & 3 & -3 & 906.7 & 0.201 & -1082845 & 3.4 & 13.2 & 9.8 & -4.0 & 7.9\\[-.3cm]
 \text{(0, 3)} & 1 & 1 & 4 & 4 & 0 & 1 & 3 & 3 & 908.9 & 0.202 & 1082924 & 3.2 & 13.2 & 10.0 & -4.0 & 7.9\\[-.3cm]
 \text{(0, 3)} & 1 & 2 & 3 & 0 & 1 & 1 & 4 & 0 & 93.5 & 0.025 & 446 & 5.8 & -3.6 & -9.4 & -7.0 & 14.1\\
 \hline \text{(1, 1)} & 1 & 2 & 1 & 1 & 0 & 1 & 0 & 0 & 948.8 & 0.641 & 1192371 & 0.4 & 0.5 & 0.0 & -4.3 & 8.6\\[-.3cm]
 \text{(1, 1)} & 1 & 2 & 1 & -1 & 0 & 1 & 1 & 0 & 950.5 & 0.113 & -1192392 & -8.1 & 0.7 & 8.8 & 64.0 & -128.0\\[-.3cm]
 \text{(1, 1)} & 1 & 2 & 1 & 0 & 0 & 1 & 0 & 0 & 947.6 & 0.784 & 3 & -1.1 & -1.1 & 0.0 & 8.6 & -17.2\\[-.3cm]
 \text{(1, 2)} & 1 & 2 & 2 & 0 & 1 & 1 & 3 & 0 & 95.9 & 0.028 & -297 & 12.5 & 2.3 & -10.1 & -31.8 & 63.5\\[-.3cm]
 \text{(1, 3)} & 1 & 2 & 3 & 2 & 1 & 1 & 4 & 3 & 92.1 & 0.026 & -121796 & -3.7 & 0.0 & 3.7 & 4.6 & -9.3\\[-.3cm]
 \text{(1, 3)} & 1 & 2 & 3 & -2 & 1 & 1 & 4 & -3 & 92.4 & 0.025 & 121798 & -3.8 & 0.0 & 3.8 & 4.6 & -9.3\\[-.3cm]
 \text{(1, 3)} & 1 & 1 & 4 & 4 & 0 & 1 & 3 & 3 & 887. & 0.204 & 1078708 & 3.6 & 14.8 & 11.3 & -4.6 & 9.2\\[-.3cm]
 \text{(1, 3)} & 1 & 1 & 4 & -4 & 0 & 1 & 3 & -3 & 884.9 & 0.202 & -1078713 & 3.8 & 14.8 & 11.0 & -4.6 & 9.2\\[-.3cm]
 \text{(1, 3)} & 1 & 2 & 3 & 0 & 1 & 1 & 4 & 0 & 92.2 & 0.025 & 315 & 6.6 & -4.0 & -10.6 & -8.2 & 16.4
 \\
 \hline \end{array}
\]
\caption{\label{table:RF transitions shifts}Systematic shifts of selected
radio-frequency M1 transitions $(v,\, L,\, F,\, S,\, J,\, J_{z})\rightarrow(v,\, L,\, F',\, S',\, J',\, J_{z}')$
(lower $\rightarrow$ upper). $f_{0}$ is the transition frequency
(excluding the quadrupole shift, including Zeeman shift for 1 Gauss).
$l,\, u$ refers to the lower and upper state, respectively. The intensity
of a transition is normalized to the strongest radio-frequency transition
having the same value of $|J_{z}-J'_{z}|$ and in the same rovibrational
level. $\Delta f_{B}$ denotes the Zeeman shift of the transition
frequency in a magnetic field of 1~G;\lyxdeleted{Schiller}{Thu Sep 12 21:20:39 2013}{
} $\Delta f_{Q}=(\Delta E_{Q})_{u}-(\Delta E_{Q})_{l}$ is the electric
quadrupole shift of the transition in a field gradient $Q_{zz}=10^{8}$~V/m\textsuperscript{2},
while $(\Delta E_{Q})_{l}$,~$(\Delta E_{Q})_{u}$ are the electric
quadrupole shifts of the lower and upper states, respectively, here
given in Hz.$\Delta\alpha^{(t)}=(\alpha^{(t)})_{u}-(\alpha^{(t)})_{l}$,~$\Delta\alpha^{(l)}=(\alpha^{(l)})_{u}-(\alpha^{(l)})_{l}$
\lyxadded{schiller}{Sun Oct 27 01:12:18 2013}{}are the transvese
and longitudinal differential electric polarisabilities betweeen upper
and lower state, respectively, in atomic units and in \lyxdeleted{schiller}{Sun Oct 27 01:09:42 2013}{zero}\lyxadded{schiller}{Sun Oct 27 01:09:46 2013}{1~G}
magnetic field. The near-zero quadrupole shift in the state $(v=1,\, L=3,\, F=1,\, S=2,\, J=3,\, J_{z}=\pm2)$
is a coincidence.}
\end{sidewaystable}

\subsection{Rotational transitions}

The two most easily accessible rotational transitions have been considered
in the search, namely the ones occurring in the ground vibrational
level $v=0$ and having the lowest transition frequencies: $(v=0,\, L=0)\rightarrow(0,\,1)$
at 1.3~THz and $(v=0,\, L=1)\rightarrow(0,\,2)$ at 2.6~THz. Of
these, the 1.3~THz transition has already been observed experimentally
\cite{Shen et al-1}. Table \ref{tab:rotational transitions} reports
selected hyperfine components. For each of the two cases $\Delta J_{z}=0$
and $\Delta J_{z}=\pm1$ those transitions having lowest absolute
Zeeman shift $|\Delta f_{B}|$ in a magnetic field of 1~G are listed.
The transitions in $(v=0,\, L=0)\rightarrow(0,\,1)$ have comparatively
large Zeeman shifts, leaving as the most interesting transitions the
``strechted-state'' doublet at 10.1~MHz, whose two components have
equal and opposite Zeeman shift and for which the electric quadrupole
shift is 7.9~Hz in a 10\textsuperscript{8}~V/m\textsuperscript{2}
field gradient. The second rotational transition listed, $(v=0,\, L=1)\rightarrow(0,\,2)$,
contains one hyperfine component with a particularly small quadratic
second-order Zeeman shift (9~Hz at 1~G) and moderate electric quadrupole
shift (-13.5~Hz). By a careful measurement of the absolute frequency
shift of this transition as a function of applied magnetic field,
it appears possible to achieve an uncertainty of the Zeeman shift
equal to 0.2\% of the value at 1~G, or approximately 0.04~Hz ($2\times10^{-14}$
relative to the absolute transition frequency),

\begin{sidewaystable}
\[
\begin{array}{|c|c|cccr|cccr|r|r|r|r|r|r|r|r|}
\hline
(\text{\textit{\ensuremath{v}}}\text{\textit{\ensuremath{'}}},\text{\textit{\ensuremath{L}}}\text{\textit{\ensuremath{'}}})
& (\text{\textit{\ensuremath{v}}},\text{\textit{\ensuremath{L}}})
& F' & S' & J' & J_{z}' & F & S & J & J_{z} & \text{freq.(1 G)} &
\text{rel.} & \Delta\text{\textit{\ensuremath{f}}}_{B}\text{(1 G)}
& \Delta\text{\textit{\ensuremath{f}}}_{Q}\text{(1 G)} &
\text{(\ensuremath{\Delta}}\text{\textit{\ensuremath{E}}}_{Q})_{u}
&
\text{(\ensuremath{\Delta}}\text{\textit{\ensuremath{E}}}_{Q})_{l}
& \Delta\alpha^{(t)} & \Delta\alpha^{(l)}\\
\text{upper} & \text{lower} & \text{} & \text{} & \text{} & \text{} & \text{} & \text{} & \text{} & \text{} & \text{[MHz]} & \text{int.} & \text{[Hz]} & \text{[Hz]} & \text{[Hz]} & \text{[Hz]} & \text{[at.u.]} & \text{[at.u.]}\\
\hline\hline \text{(0, 1)} & \text{(0, 0)} & 0 & 1 & 1 & 0 & 0 & 1 & 1 & 0 & 1.7 & 0.002 & -867 & 7.8 & 7.8 & 0 & -449.7 & -274.5\\
\text{(0, 1)} & \text{(0, 0)} & 1 & 2 & 1 & 0 & 1 & 2 & 2 & 0 & -33.2 & 0.42 & -2780 & -0.9 & -0.9 & 0 & -384.2 & -405.4\\
\text{(0, 1)} & \text{(0, 0)} & 0 & 1 & 2 & 0 & 0 & 1 & 1 & 0 & -2.1 & 0.755 & -2915 & -7.8 & -7.8 & 0 & -332.9 & -508.2\\
\text{(0, 1)} & \text{(0, 0)} & 0 & 1 & 0 & 0 & 0 & 1 & 1 & 0 & 6.1 & 0.377 & 3818 & 0.0 & 0.0 & 0 & -391.3 & -391.3\\
\text{(0, 1)} & \text{(0, 0)} & 1 & 0 & 1 & 0 & 1 & 0 & 0 & 0 & -9.1 & 1. & 5050 & -13.1 & -13.1 & 0 & -293.7 & -586.5\\
\text{(0, 1)} & \text{(0, 0)} & 1 & 1 & 2 & 0 & 1 & 1 & 1 & 0 & 11.8 & 0.756 & -6171 & -7.9 & -7.9 & 0 & -332.8 & -508.3\\
\text{(0, 1)} & \text{(0, 0)} & 1 & 2 & 3 & \text{\ensuremath{\pm}3} & 1 & 2 & 2 & \text{\ensuremath{\pm}2} & 10.1 & 1. & \text{\ensuremath{\mp}558} & 7.9 & 7.9 & 0 & -449.8 & -274.3\\
\hline \text{(0, 2)} & \text{(0, 1)} & 0 & 1 & 2 & 0 & 0 & 1 & 1 & 0 & 0.2 & 0.798 & 9 & -13.5 & -5.6 & 7.8 & 72.2 & -144.4\\
\text{(0, 2)} & \text{(0, 1)} & 0 & 1 & 3 & 0 & 0 & 1 & 2 & 0 & -2.1 & 0.957 & 792 & -1.2 & -9.0 & -7.8 & -36.3 & 72.7\\
\text{(0, 2)} & \text{(0, 1)} & 0 & 1 & 1 & 0 & 0 & 1 & 0 & 0 & 1.8 & 0.886 & -840 & -7.9 & -7.9 & 0.0 & 19.3 & -38.6\\
\text{(0, 2)} & \text{(0, 1)} & 1 & 2 & 4 & \text{\ensuremath{\pm}4} & 1 & 2 & 3 & \text{\ensuremath{\pm}3} & 12.9 & 1. & \text{\ensuremath{\mp}558} & 3.4 & 11.3 & 7.9 & 30.9 & -61.7
\\\hline \end{array}
\]
\caption{\label{tab:rotational transitions} Systematic shifts of selected
rotational transitions in the vibrational ground state $v=0$. An
entry having two signs for $J_{z}$ and $J'_{z}$ indicates the two
transitions between streched states. The frequency value is the spin-dependent
contribution to the total transition frequency $f_{0}$. For the $(0,0)\rightarrow(0,1)$
transition, $f_{0}\simeq$1.3~THz, For the $(0,1)\rightarrow(0,2)$
transition, $f_{0}\simeq$2.6~THz, The intensity of each transition
is normalized to that of the strongest transition of the particular
rotational transition having the same $|\Delta J_{z}$|. Other notations
are as in Table~\ref{table:RF transitions shifts}. }
\end{sidewaystable}

\subsection{Rovibrational transitions}

\begin{sidewaystable}
\[
 \begin{array}{|c|c|cccr|cccr|r|r|r|r|r|r|r|r|}
 \hline
 (\text{\textit{\ensuremath{v}}}\textit{'},\text{\textit{\ensuremath{L}}}\text{'})
 & (\text{\textit{\ensuremath{v}}},\text{\textit{\ensuremath{L}}}) & F' & S' & J'
 & J_{z}' & F & S & J & J_{z} & \text{freq.(1 G)} & \text{rel.}
 & \Delta\text{\textit{\ensuremath{f}}}_{B}\text{(1 G)}
 & \Delta\text{\textit{\ensuremath{f}}}_{Q}\text{(1 G)}
 & \text{(\ensuremath{\Delta}}\text{\textit{\ensuremath{E}}}_{Q})_{u}
 & \text{(\ensuremath{\Delta}}\text{\textit{\ensuremath{E}}}_{Q})_{l}
 & \Delta\alpha^{(t)} & \Delta\alpha^{(l)}\\
 \text{upper} & \text{lower} & \text{} & \text{} & \text{} & \text{}
 & \text{} & \text{} & \text{} & \text{} & \text{[MHz]} & \text{int.}
 & \text{[Hz]} & \text{[Hz]} & \text{[Hz]} & \text{[Hz]} & \text{[at.u.]}
 & \text{[at.u.]}\\
 \hline\hline
 \text{(1, 5)} & \text{(0, 4)} & 0 & 1 & 5 & 0 & 0 & 1 & 4 & 0 & 16.8 & 0.97 & 29.5 & -1.6 & -10.5 & -8.8 & -0.5 & 3.4\\[-.3cm]
 \text{(1, 5)} & \text{(0, 4)} & 1 & 2 & 5 & 0 & 1 & 2 & 4 & 0 & -3.1 & 0.9 & -57.3 & -2.3 & -8.5 & -6.2 & 0.4 & 1.6\\
 \hline
 \text{(2, 4)} & \text{(0, 3)} & 0 & 1 & 4 & 0 & 0 & 1 & 3 & 0 & 31.8 & 0.96 & 24.0 & -3.1 & -11.1 & -7.9 & 0.5 & 4.0\\[-.3cm]
 \text{(2, 4)} & \text{(0, 3)} & 0 & 1 & 5 & 0 & 0 & 1 & 4 & 0 & 30.7 & 0.99 & 31.7 & -2.7 & -12.2 & -9.5 & -0.5 & 6.0\\[-.3cm]
 \text{(2, 5)} & \text{(0, 4)} & 0 & 1 & 5 & 0 & 0 & 1 & 4 & 0 & 32.0 & 0.97 & -38.9 & -2.9 & -11.7 & -8.8 & 1.1 & 2.8\\[-.3cm]
 \text{(2, 5)} & \text{(0, 4)} & 0 & 1 & 6 & 0 & 0 & 1 & 5 & 0 & 31.2 & 1 & -39.6 & -2.7 & -12.4 & -9.7 & 0.9 & 3.4\\
 \hline
 \text{(3, 2)} & \text{(0, 1)} & 1 & 1 & 3 & 0 & 1 & 1 & 2 & 0 & -3.8 & 0.95 & 21.2 & -4.7 & -12.6 & -7.9 & -20.8 & 49.6\\[-.3cm]
 \text{(3, 3)} & \text{(0, 2)} & 0 & 1 & 4 & 0 & 0 & 1 & 3 & 0 & 44.7 & 0.98 & -12.4 & -4.2 & -13.2 & -9. & -1.5 & 11.1\\[-.3cm]
 \text{(3, 3)} & \text{(0, 2)} & 1 & 0 & 3 & 0 & 1 & 0 & 2 & 0 & -10.8 & 1. & 54.8 & -3.8 & -13.9 & -10.1 & -3.3 & 14.7\\[-.3cm]
 \text{(3, 4)} & \text{(0, 3)} & 1 & 2 & 4 & 0 & 1 & 2 & 3 & 0 & -16.0 & 0.84 & 6.4 & -5.0 & -8.6 & -3.6 & 5.2 & -2.0\\[-.3cm]
 \text{(3, 4)} & \text{(0, 3)} & 1 & 1 & 5 & 0 & 1 & 1 & 4 & 0 & -8.4 & 0.99 & -11.2 & -4.1 & -13.5 & -9.4 & 2.0 & 4.3\\[-.3cm]
 \text{(3, 4)} & \text{(0, 3)} & 0 & 1 & 4 & 0 & 0 & 1 & 3 & 0 & 45.9 & 0.96 & -55.0 & -4.4 & -12.3 & -7.9 & 2.9 & 2.5\\
 \hline
 \text{(4, 2)} & \text{(0, 1)} & 0 & 1 & 1 & 0 & 0 & 1 & 0 & 0 & 59.1 & 0.87 & -17.7 & -12.2 & -12.2 & 0.0 & 39.8 & -67.5\\[-.3cm]
 \text{(4, 2)} & \text{(0, 1)} & 0 & 1 & 3 & 0 & 0 & 1 & 2 & 0 & 57.5 & 0.94 & -36.7 & -6.1 & -14. & -7.8 & -13.5 & 39.1\\[-.3cm]
 \text{(4, 3)} & \text{(0, 2)} & 0 & 1 & 3 & 0 & 0 & 1 & 2 & 0 & 58.9 & 0.91 & -37.2 & -6.7 & -12.3 & -5.6 & 7.7 & -3.2\\[-.3cm]
 \text{(4, 5)} & \text{(0, 4)} & 1 & 1 & 4 & 0 & 1 & 1 & 3 & 0 & -11.4 & 0.98 & -21.1 & -5.6 & -14.8 & -9.1 & 5.4 & 1.8\\[-.3cm]
 \text{(4, 5)} & \text{(0, 4)} & 1 & 2 & 3 & 0 & 1 & 2 & 2 & 0 & -16.1 & 0.96 & -29.0 & -5.7 & -13.9 & -8.2 & 5.6 & 1.3\\[-.3cm]
 \text{(4, 5)} & \text{(0, 4)} & 1 & 2 & 7 & 0 & 1 & 2 & 6 & 0 & -24.8 & 0.98 & -30.8 & -5.5 & -14.8 & -9.3 & 5.3 & 1.9\\[-.3cm]
 \text{(4, 5)} & \text{(0, 4)} & 1 & 1 & 5 & 0 & 1 & 1 & 4 & 0 & -11.9 & 0.98 & -34.3 & -5.6 & -14.7 & -9.2 & 5.4 & 1.9\\[-.3cm]
 \text{(4, 5)} & \text{(0, 4)} & 0 & 1 & 4 & 0 & 0 & 1 & 3 & 0 & 59.0 & 0.99 & -52.6 & -5.6 & -15.1 & -9.5 & 5.3 & 1.9\\
 \hline
 \text{(5, 4)} & \text{(0, 3)} & 0 & 1 & 3 & 0 & 0 & 1 & 2 & 0 & 71.1 & 0.98 & -2.3 & -7.2 & -16.3 & -9.1 & 8.9 & -0.2\\[-.3cm]
 \text{(5, 5)} & \text{(0, 4)} & 0 & 1 & 4 & 0 & 0 & 1 & 3 & 0 & 70.8 & 0.99 & 45.9 & -7.2 & -16.7 & -9.5 & 8.3 & 1.4
 \\
 \hline \end{array}
 \]

\caption{\label{tab: rovibrational transitions} Selected
rovibrational transitions with small quadratic Zeeman shifts at
1~G. For the $(0,3)\rightarrow(2,4)$ transition, the absolute
frequency $f_{0}\simeq$116~THz, for the $(0,2)\rightarrow(3,3)$
transition, $f_{0}\simeq$166~THz, \lyxdeleted{schiller}{Sat Oct 26
13:06:00 2013}{and }for the $(0,1)\rightarrow(4,2)$ transition,
$f_{0}\simeq$214~THz\lyxadded{schiller}{Sat Oct 26 14:02:07
2013}{, and for the $(0,3)\rightarrow(5,4)$ transition,
$f_{0}\simeq$261~THz}. See caption of Tab.~\ref{tab:rotational
transitions} for explanations.}
\end{sidewaystable}

The search for favorable rovibrational transitions was limited to
transitions originating in $v=0,\,1$ and ending in $v'\le5$. A
subset of transitions was selected according to the criterium that
their Zeeman shifts are less than 60~Hz for fields less than 1~G.
 The transitions
originating from $v=1$ do not offer any advantages compared to
those originating from the ground vibrational state, and we limit
the following discussion to the latter. They are shown in
Tab.~\ref{tab: rovibrational transitions}. These are all
$J_{z}=0\rightarrow J'_{z}=0$ transitions. Two transitions (at
-16.0 and 71.1~MHz) have particularly low Zeeman shifts, 6 and
-2~Hz at 1~G, respectively.
The small differential Zeeman shifts do not arise from strong
cancellation of large individual shifts, but from cancellation of
moderate shifts: For example, the -2.3 Hz shift results from
individual shifts of 58 Hz and 60 Hz, while the 6.3 Hz shift from
two individual shifts of approximately 6.2 kHz. The latter
represents the largest relative cancellation of all transitions in
Table 7, and is still consistent with the nonrelativistic
approximations inherent in the Zeeman shift calculation.
If it is possible to minimize the magnetic field in the trap, e.g.
to 0.02~G, the quadratic Zeeman shift is reduced by a factor
$\simeq2500$, to a relative level of approx. $2\times10^{-17}$ and
$4\times10^{-18}$, respectively. This is a negligible shift,
compared to the other systematic effects discussed here. The
electric quadrupole shift of these transitions is approx.
$-3\times10^{-14}$ at the given gradient value.

Another transition worth noting is the
$(1,\,0,\,1,\,0)\rightarrow(1,\,0,\,2,\,0)$ transition in
$(v=0,\,L=1)\rightarrow(2,\,2)$ (not shown in Table 7), which has
$\Delta f_B=102.5\,$Hz at 1 G, and one of the lowest fractional
electric quadrupole shifts, $\Delta f_Q=0.26\,$Hz
($2.3\times10^{-15}$). The relatively large Zeeman shift at 1 G
would be reduced to the $4\times10^{-16}$ level in a 0.02 G field.

If, however, the magnetic field is at the 1~G level, for which the
Zeeman shift is appreciable, one may determine the shift precisely
by measuring the frequency shift of the transition as a function
of applied magnetic field. Suppose that \lyxadded{schiller}{Sat
Oct 26 13:59:23 2013}{the}\lyxdeleted{schiller}{Sat Oct 26
13:59:23 2013}{a} transition frequency can be measured with a
resolution equal to 1\% of the natural
linewidth\lyxadded{schiller}{Sat Oct 26 13:59:39 2013}{ at each
magnetic field value}, e.g. 0.14~Hz for a transition
$v=0\rightarrow v'=3$. \lyxdeleted{schiller}{Sat Oct 26 13:14:01
2013}{}\lyxadded{schiller}{Sat Oct 26 13:59:58 2013}{ The result
of the Zeeman shift evaluation may then }reach an uncertainty of
0.04~Hz, or $2\times10^{-16}$ relative to the absolute transition
frequency\lyxadded{schiller}{Sat Oct 26 23:21:32 2013}{ of this
overtone transition}.

A second set of transitions are the stretched-states doublets,
tabulated in Tab.~\ref{tab: rovibrational stretched-state
transitions}. \lyxadded{schiller}{Sat Oct 26 13:21:20 2013}{For
space reasons, we have not included transitions to $v'=5$ or
$L'=5$ levels. }Their linear Zeeman shift is approximately
$\pm0.5$~kHz/G. Suppose that each transition frequency of a
doublet can be measured with a resolution equal to 1\% of the
natural linewidth, e.g. 0.14~Hz for the transition $v=0\rightarrow
v=3$. Then the Zeeman effect uncertainty of the mean of the
doublet frequencies would be 0.2~Hz, or approximately
$1\times10^{-15}$ relative to the absolute transition frequency.
Repeating this for a set of magnetic field values could reduce the
error to $2\times10^{-16}$. The electric quadrupole shift of this
particular transition is one order of magnitude smaller than the
typical shift of all other stretched state transitions, $-0.3$~Hz
versus several Hz, or $2\times10^{-15}$ relative to the absolute
transition frequency.

 \begin{sidewaystable*}
 \noindent \centering{}$\begin{array}{|c|c|cccr|cccr|c|r|r|r|r|r|r|r|}
 \hline
 (\text{\textit{\ensuremath{v}}}\textit{\ensuremath{'}},\text{\textit{\ensuremath{L}}}\textit{\ensuremath{'}})
 & (\text{\textit{\ensuremath{v}}},\text{\textit{\ensuremath{L}}}) & F' & S' & J' & J_{z}' & F & S & J &
 J_{z}
 & \text{freq.(1 G)} & \text{rel.} & \Delta\text{\textit{\ensuremath{f}}}_{B}\text{(1 G)}
 & \Delta\text{\textit{\ensuremath{f}}}_{Q}\text{(1 G)}
 & \text{(\ensuremath{\Delta}}\text{\textit{\ensuremath{E}}}_{Q})_{u}
 & \text{(\ensuremath{\Delta}}\text{\textit{\ensuremath{E}}}_{Q})_{l}
 & \Delta\alpha^{(t)} & \Delta\alpha^{(l)}\\[-.3cm]
 \text{upper} & \text{lower} & \text{} & \text{} & \text{} & \text{}
 & \text{} & \text{} & \text{} & \text{} & \text{[MHz]} & \text{int.}
 & \text{[Hz]} & [Hz] & \text{[Hz]} & \text{[Hz]} & \text{[at.u.]} & \text{[at.u.]}\\
 \hline\hline
 \text{(1, 0)} & \text{(0, 1)} & 1 & 2 & 2 & \text{\ensuremath{\pm}2} & 1 & 2 & 3 & \text{\ensuremath{\pm}3} & -17.0 & 1 & \text{\ensuremath{\pm}558.3} & -7.9 & 0 & 7.9 & 517.2 & 341.7\\[-.3cm]
 \text{(1, 1)} & \text{(0, 0)} & 1 & 2 & 3 & \text{\ensuremath{\pm}3} & 1 & 2 & 2 & \text{\ensuremath{\pm}2} & 2.6 & 1 & \text{\ensuremath{\mp}553.7} & 8.9 & 8.8 & 0 & -459. & -253.8\\[-.3cm]
 \text{(1, 1)} & \text{(0, 2)} & 1 & 2 & 3 & \text{\ensuremath{\pm}3} & 1 & 2 & 4 & \text{\ensuremath{\pm}4} & -20.4 & 1 & \text{\ensuremath{\pm}562.6} & -2.4 & 8.8 & 11.3 & -40.1 & 82.2\\[-.3cm]
 \text{(1, 2)} & \text{(0, 1)} & 1 & 2 & 4 & \text{\ensuremath{\pm}4} & 1 & 2 & 3 & \text{\ensuremath{\pm}3} & 4.8 & 1 & \text{\ensuremath{\mp}548.9} & 4.8 & 12.7 & 7.9 & 26.9 & -51.7\\[-.3cm]
 \text{(1, 2)} & \text{(0, 3)} & 1 & 2 & 4 & \text{\ensuremath{\pm}4} & 1 & 2 & 5 & \text{\ensuremath{\pm}5} & -21.6 & 1 & \text{\ensuremath{\pm}566.6} & -0.6 & 12.7 & 13.2 & -15.6 & 33.4\\[-.3cm]
 \text{(1, 3)} & \text{(0, 2)} & 1 & 2 & 5 & \text{\ensuremath{\pm}5} & 1 & 2 & 4 & \text{\ensuremath{\pm}4} & 4.6 & 1 & \text{\ensuremath{\mp}543.8} & 3.6 & 14.9 & 11.3 & 9.8 & -17.4\\[-.3cm]
 \text{(1, 3)} & \text{(0, 4)} & 1 & 2 & 5 & \text{\ensuremath{\pm}5} & 1 & 2 & 6 & \text{\ensuremath{\pm}6} & -22.3 & 1 & \text{\ensuremath{\pm}570.5} & 0.3 & 14.9 & 14.6 & -7.6 & 17.3\\[-.3cm]
 \text{(1, 4)} & \text{(0, 3)} & 1 & 2 & 6 & \text{\ensuremath{\pm}6} & 1 & 2 & 5 & \text{\ensuremath{\pm}5} & 3.8 & 1 & \text{\ensuremath{\mp}538.4} & 3.1 & 16.3 & 13.2 & 4.8 & -7.3\\
 \hline
 \text{(2, 0)} & \text{(0, 1)} & 1 & 2 & 2 & \text{\ensuremath{\pm}2} & 1 & 2 & 3 & \text{\ensuremath{\pm}3} & -23.5 & 1 & \text{\ensuremath{\pm}558.3} & -7.9 & 0 & 7.9 & 595.2 & 419.7\\[-.3cm]
 \text{(2, 1)} & \text{(0, 0)} & 1 & 2 & 3 & \text{\ensuremath{\pm}3} & 1 & 2 & 2 & \text{\ensuremath{\pm}2} & -4.4 & 1 & \text{\ensuremath{\mp}548.9} & 9.9 & 9.9 & 0 & -469.6 & -230.0\\[-.3cm]
 \text{(2, 1)} & \text{(0, 2)} & 1 & 2 & 3 & \text{\ensuremath{\pm}3} & 1 & 2 & 4 & \text{\ensuremath{\pm}4} & -27.4 & 1 & \text{\ensuremath{\pm}567.4} & -1.4 & 9.9 & 11.3 & -50.6 & 106.0\\[-.3cm]
 \text{(2, 2)} & \text{(0, 1)} & 1 & 2 & 4 & \text{\ensuremath{\pm}4} & 1 & 2 & 3 & \text{\ensuremath{\pm}3} & -2.9 & 1 & \text{\ensuremath{\mp}539.1} & 6.3 & 14.2 & 7.9 & 22.5 & -40.1\\[-.3cm]
 \text{(2, 2)} & \text{(0, 3)} & 1 & 2 & 4 & \text{\ensuremath{\pm}4} & 1 & 2 & 5 & \text{\ensuremath{\pm}5} & -29.3 & 1 & \text{\ensuremath{\pm}576.4} & 0.9 & 14.2 & 13.2 & -20.1 & 44.9\\[-.3cm]
 \text{(2, 3)} & \text{(0, 2)} & 1 & 2 & 5 & \text{\ensuremath{\pm}5} & 1 & 2 & 4 & \text{\ensuremath{\pm}4} & -3.7 & 1 & \text{\ensuremath{\mp}529.1} & 5.3 & 16.6 & 11.3 & 7.7 & -10.5\\[-.3cm]
 \text{(2, 3)} & \text{(0, 4)} & 1 & 2 & 5 & \text{\ensuremath{\pm}5} & 1 & 2 & 6 & \text{\ensuremath{\pm}6} & -30.6 & 1 & \text{\ensuremath{\pm}585.2} & 2.1 & 16.6 & 14.6 & -9.8 & 24.2\\[-.3cm]
 \text{(2, 4)} & \text{(0, 3)} & 1 & 2 & 6 & \text{\ensuremath{\pm}6} & 1 & 2 & 5 & \text{\ensuremath{\pm}5} & -5.2 & 1 & \text{\ensuremath{\mp}518.8} & 5.0 & 18.3 & 13.2 & 3.8 & -2.6\\
 \hline
 \text{(3, 0)} & \text{(0, 1)} & 1 & 2 & 2 & \text{\ensuremath{\pm}2} & 1 & 2 & 3 & \text{\ensuremath{\pm}3} & -29.5 & 1 & \text{\ensuremath{\pm}558.3} & -7.9 & 0 & 7.9 & 685.9 & 510.4\\[-.3cm]
 \text{(3, 1)} & \text{(0, 0)} & 1 & 2 & 3 & \text{\ensuremath{\pm}3} & 1 & 2 & 2 & \text{\ensuremath{\pm}2} & -10.9 & 1 & \text{\ensuremath{\mp}543.7} & 11.0 & 11.0 & 0 & -481.8 & -202.4\\[-.3cm]
 \text{(3, 1)} & \text{(0, 2)} & 1 & 2 & 3 & \text{\ensuremath{\pm}3} & 1 & 2 & 4 & \text{\ensuremath{\pm}4} & -33.9 & 1 & \text{\ensuremath{\pm}572.6} & -0.3 & 11.0 & 11.3 & -62.9 & 133.6\\[-.3cm]
 \text{(3, 2)} & \text{(0, 1)} & 1 & 2 & 4 & \text{\ensuremath{\pm}4} & 1 & 2 & 3 & \text{\ensuremath{\pm}3} & -10.1 & 1 & \text{\ensuremath{\mp}528.7} & 7.9 & 15.8 & 7.9 & 17.4 & -26.7\\[-.3cm]
 \text{(3, 2)} & \text{(0, 3)} & 1 & 2 & 4 & \text{\ensuremath{\pm}4} & 1 & 2 & 5 & \text{\ensuremath{\pm}5} & -36.5 & 1 & \text{\ensuremath{\pm}586.8} & 2.5 & 15.8 & 13.2 & -25.2 & 58.4\\[-.3cm]
 \text{(3, 3)} & \text{(0, 2)} & 1 & 2 & 5 & \text{\ensuremath{\pm}5} & 1 & 2 & 4 & \text{\ensuremath{\pm}4} & -11.6 & 1 & \text{\ensuremath{\mp}513.4} & 7.2 & 18.5 & 11.3 & 5.3 & -2.4\\[-.3cm]
 \text{(3, 3)} & \text{(0, 4)} & 1 & 2 & 5 & \text{\ensuremath{\pm}5} & 1 & 2 & 6 & \text{\ensuremath{\pm}6} & -38.5 & 1 & \text{\ensuremath{\pm}600.9} & 3.9 & 18.5 & 14.6 & -12.1 & 32.2\\[-.3cm]
 \text{(3, 4)} & \text{(0, 3)} & 1 & 2 & 6 & \text{\ensuremath{\pm}6} & 1 & 2 & 5 & \text{\ensuremath{\pm}5} & -26.2 & 1 & \text{\ensuremath{\mp}497.7} & 7.1 & 20.3 & 13.2 & 2.7 & 2.8\\
 \hline
 \text{(4, 0)} & \text{(0, 1)} & 1 & 2 & 2 & \text{\ensuremath{\pm}2} & 1 & 2 & 3 & \text{\ensuremath{\pm}3} & -35.2 & 1 & \text{\ensuremath{\pm}558.3} & -7.9 & 0 & 7.9 & 791.8 & 616.3\\[-.3cm]
 \text{(4, 1)} & \text{(0, 0)} & 1 & 2 & 3 & \text{\ensuremath{\pm}3} & 1 & 2 & 2 & \text{\ensuremath{\pm}2} & -17.0 & 1 & \text{\ensuremath{\mp}538.1} & 12.2 & 12.2 & 0 & -496.0 & -170.0\\[-.3cm]
 \text{(4, 1)} & \text{(0, 2)} & 1 & 2 & 3 & \text{\ensuremath{\pm}3} & 1 & 2 & 4 & \text{\ensuremath{\pm}4} & -40.0 & 1 & \text{\ensuremath{\pm}578.1} & 0.9 & 12.2 & 11.3 & -77.1 & 165.9\\[-.3cm]
 \text{(4, 2)} & \text{(0, 1)} & 1 & 2 & 4 & \text{\ensuremath{\pm}4} & 1 & 2 & 3 & \text{\ensuremath{\pm}3} & -16.8 & 1 & \text{\ensuremath{\mp}517.6} & 9.6 & 17.5 & 7.9 & 11.5 & -11.\\[-.3cm]
 \text{(4, 2)} & \text{(0, 3)} & 1 & 2 & 4 & \text{\ensuremath{\pm}4} & 1 & 2 & 5 & \text{\ensuremath{\pm}5} & -43.2 & 1 & \text{\ensuremath{\pm}597.9} & 4.2 & 17.5 & 13.2 & -31.1 & 74.1\\[-.3cm]
 \text{(4, 3)} & \text{(0, 2)} & 1 & 2 & 5 & \text{\ensuremath{\pm}5} & 1 & 2 & 4 & \text{\ensuremath{\pm}4} & -18.9 & 1 & \text{\ensuremath{\mp}496.6} & 9.2 & 20.5 & 11.3 & 2.6 & 7.0\\[-.3cm]
 \text{(4, 3)} & \text{(0, 4)} & 1 & 2 & 5 & \text{\ensuremath{\pm}5} & 1 & 2 & 6 & \text{\ensuremath{\pm}6} & -45.8 & 1 & \text{\ensuremath{\pm}617.7} & 5.9 & 20.5 & 14.6 & -14.8 & 41.6\\[-.3cm]
 \text{(4, 4)} & \text{(0, 3)} & 1 & 2 & 6 & \text{\ensuremath{\pm}6} & 1 & 2 & 5 & \text{\ensuremath{\pm}5} & -21.7 & 1 & \text{\ensuremath{\mp}475.2} & 9.3 & 22.5 & 13.2 & 1.6 & 9.1
 \\
 \hline
 \end{array}$\caption{\label{tab: rovibrational stretched-state transitions} Rovibrational
transitions between stretched hyperfine states. The double sign refers
to the pair of transitions $J_{z}=J\rightarrow\, J'_{z}=J'$ and $J_{z}=-J\rightarrow\, J'_{z}=-J'$
, which have opposite Zeeman shifts, but the same electric quadrupole
shift. The absolute transition frequencies are similar to those of
table \ref{tab: rovibrational transitions}. See caption of Tab.~\ref{tab:rotational transitions}
for explanations.}
\end{sidewaystable*}

\subsection{Two-photon rovibrational transitions}

Two-photon transitions (E2) are of interest since they can be
excited with suppression of first-order Doppler shift even without
strong spatial confinement of the ions. These transitions were
discussed for $\HDp$ in Ref.~\cite{older}. It was subsequently
shown in Ref.~\cite{Zee1} that there exist two-photon transitions
without any Zeeman shift as well as stretched-state transitions.
Tab.~\ref{tab:two-photon transitions} reports two-photon
transitions between levels having low values of $L$ and $L'$.
These are favourable from the experimental point of view, since
the number of two-photon transitions arising from a pair of levels
is reduced when the angular momenta are small, which translates in
a higher transition strength per transition. Also, the ease of
populating sufficiently strongly the lower hyperfine states is
simplified.

The most favorable transition from the point of view of the
systematic shifts due to magnetic field and electric field
gradient is the stretched-state transition of $(v=0,\,
L=0)\rightarrow(v'=2,\, L'=0)$, as both effects are absent. The
stretched-state transition of $(0,\,1)\rightarrow(2,\,1)$ is also
advantageous. For the latter, assuming the same criterium as
above, the Zeeman effect uncertainty would be 0.03~Hz, or
approximately $3\times10^{-16}$ relative to the absolute
two-photon transition frequency. The electric quadrupole shift is
2~Hz, nearly two orders larger.

\begin{sidewaystable}
\centering{}$\begin{array}{|c|c|cccc|cccc|c|r|r|r|r|r|r|r|} \hline
(\text{\textit{\ensuremath{v'}}},\text{\textit{\ensuremath{L'}}})
& (\text{\textit{\ensuremath{v}}},\text{\textit{\ensuremath{L}}})
& F' & S' & J' & J'_{z} & F & S & J & J_{z} & \text{freq.(1\,\ G)}
& \text{rel.} &
\Delta\text{\textit{\ensuremath{f}}}_{B}\text{(1\,G)} &
\Delta\text{\textit{\ensuremath{f}}}_{Q} \text{(1\,G)}&
\text{(\ensuremath{\Delta}}\text{\textit{\ensuremath{E}}}_{Q})_{u}
&
\text{(\ensuremath{\Delta}}\text{\textit{\ensuremath{E}}}_{Q})_{l}
& \Delta\alpha^{(t)} & \Delta\alpha^{(l)}\\
\text{upper} & \text{lower} & \text{} & \text{} & \text{} &
\text{} & \text{} & \text{} & \text{} & \text{} & \text{[MHz]} &
\text{int.} & \text{[Hz]} & \text{[Hz]} & \text{[Hz]} &
\text{[Hz]} & \text{[at.u.]} & \text{[at.u.]}\\
\hline\hline \text{\ensuremath{(2,0)}} & \text{\ensuremath{(0,0)}}
& 1 & 2 & 2 & \text{\ensuremath{\pm}2} & 1 & 2 & 2 &
\text{\ensuremath{\pm}2} & -13.4 & 1 & 0 & 0 & 0 & 0 & 145.4 &
145.4\\
\text{\ensuremath{(2,2)}} & \text{\ensuremath{(0,0)}} & 1 & 2 & 4
& \text{\ensuremath{\pm}4} & 1 & 2 & 2 & \text{\ensuremath{\pm}2}
& 7.2 & 1 & \text{\ensuremath{\mp}\ensuremath{1096}} & 14.2 & 14.2
& 0 & -427.3 & -314.4\\ \text{\ensuremath{(2,1)}} &
\text{\ensuremath{(0,1)}} & 1 & 2 & 3 & \text{\ensuremath{\pm}3} &
1 & 2 & 3 & \text{\ensuremath{\pm}3} & -14.4 & 1 &
\text{\ensuremath{\pm}\ensuremath{10}} & 2.0 & 9.9 & 7.9 & -19.8 &
44.3
\\\hline \end{array}$\caption{\label{tab:two-photon transitions} Selected two-photon transitions
with favourably low Zeeman shifts. Each line is a stetched-state doublet.
The absolute transition frequencies are $f_{0}\simeq$112,~112, 115~THz,
respectively. See caption of Tab.~\ref{tab:rotational transitions}
for explanations. }
\end{sidewaystable}

\section{Discussion}

\subsection{Quadrupole shift measurement and cancellation\label{sub:Quadrupole-shift-measurement}}

The previous section has shown that among the rovibrational transitions
having small Zeeman shifts (Tab.~\ref{tab: rovibrational transitions},~\ref{tab: rovibrational stretched-state transitions}),
the electric quadrupole shifts range from absolute values of zero
to approximately 10~Hz, in a typical gradient of $10^{8}$~V/m\textsuperscript{2}.
For those transitions for which the shift is finite (i.e. excluding
the particular two-photon transitions), the relative values range
from $\simeq1\times10^{-15}$ to the largest values $\simeq1\times10^{-13}$
in relative units. It is useful to compare these magnitudes with the
value for atomic ions used in ion optical clocks. For example, in
the mercury ion, the shift is on the order of 10~Hz for the same
gradient strength, or $1\times10^{-14}$ in relative units \cite{Itano 2000}.

Although small for selected transitions of $\HDp$, the quadrupole
shift can actually be determined and nulled. The property that the
electric quadrupole shift depends only on the componenent of the gradient
tensor in the direction of the magnetic field, allows for a determination
and cancellation of the quadrupole shift. The approach is similar
to one of the methods of quadrupole shift control applied to atomic
ions in ion optical clocks, introduced by Itano \cite{Itano 2000}.

Consider applying the magnetic field in turn along three orthogonal
spatial directions $x,\, y,\, z$, and measuring the corresponding
transition frequencies $f_{x},\, f_{y},\, f_{z}$, keeping the magnetic
field strength constant. Since $f_{i}=f_{0}+(\Delta f_{Q})_{i},\, i=x,y,z,$
and the transition frequency shift is linear in the gradient strength,
$(\Delta f_{Q})_{i}=p\, Q_{ii}$, where $p=p(v,\, L,\, n,\, J_{z},\, v',\, L',\, n',\, J'_{z})$
is the sensitivity of the particular transition frequency, we have

\begin{eqnarray}
f_{x} & = & f_{0}+p\, Q_{xx}\,,\nonumber \\
f_{y} & = & f_{0}+p\, Q_{yy}\,,\label{eq:three directions}\\
f_{z} & = & f_{0}+p\, Q_{zz}\,.\nonumber
\end{eqnarray}
Since the gradients satisfy the Laplace equation $Q_{xx}+Q_{yy}+Q_{zz}=0$,
we obtain

\begin{eqnarray}
f_{0} & = & \frac{1}{3}\,(f_{x}+f_{y}+f_{z})\,,\nonumber \\
p\, Q_{xx} & = & \frac{1}{3}(2\, f_{x}-f_{y}-f_{z})\,,\label{eq:compensation and gradients}\\
p\, Q_{yy} & = & \frac{1}{3}(2\, f_{y}-f_{x}-f_{z}).\nonumber
\end{eqnarray}
The unperturbed transition frequency $f_{0}$ is calculated from a
simple average over three directions. The error in determining it
arises from (i) the uncertainy of each measurement $f_{x},\, f_{y},\, f_{z}$,
and (ii) the inaccuracy in establishing three perfectly orthogonal
magnetic field directions and thus obtaining a perfect cancellation
of the quadrupole shift.

The first uncertainty may be estimated as previously by the 1\% assumption,
giving 0.14~Hz$/\sqrt{3}$ for rovibrational transitions $v=0\rightarrow3$.
The second uncertainty, in a precision experiment on the mercury ion
clock, was less than $5\times10^{-17}$~\cite{Oskay}. We may expect
that it will eventually be possible to achieve an equivalent uncertainty
of this type also for $\HDp$, that is, in the range between $5\times10^{-18}$
and $5\times10^{-16}$, depending on the transition (rescaling by
the sensitivity of $\HDp$ compared to Hg\textsuperscript{+}). Then,
the electric quadrupole shift nulling uncertainty for all considered
E1 rovibrational transitions will be dominated by the type-(i)-uncertainty.
With the 1\% criterium used here, this uncertainy would be approximately
$0.5$ to $1\times10^{-15}$, limited by the natural lifetime of the
upper level.

Note that since $p$ is known, the gradient strengths can also be
determined experimentally, via Eqs.~(\ref{eq:compensation and gradients}).

\subsection{Other systematic effects}

The discussion has so far concentrated on the electric quadrupole
shift and the Zeeman shift in a time-independent (d.c.) magnetic field.
Other systematic effects affecting transition frequencies of trapped
ions are the 2\textsuperscript{nd}-order Doppler shift, the Zeeman
shift due to a.c. magnetic fields of the trap, the light shifts, the
black-body radiation shift and the quadratic Stark shift due to stray
electric fields of the trap. We comment only on the latter two, since
we believe that the others are negligible, with the possible exception
of the light shift in case of two-photon transitions. The black-body
radiation shift at 300~K is of order $1\times10^{-16}$ for the transitions
discussed here \cite{kool}. By an accurate determination of the environment
temperature or by use of a cryogenic ion trap the uncertainty of this
shift can be reduced further by at least one order.

The Stark frequncy shift of a transition frequency is given
by\lyxadded{schiller}{Sun Oct 27 01:16:22 2013}{} $\Delta
E_{S}=-(\Delta\alpha^{(l)}\,
E_{z}^{2}+\Delta\alpha^{(t)}\,(E_{x}^{2}+E_{y}^{2}))/2$, where
$E_{x},\, E_{y},\, E_{z}$ are the components of the electric
field, and\lyxadded{schiller}{Sun Oct 27 01:16:26 2013}{}
$\Delta\alpha^{(l)}$, $\Delta\alpha^{(t)}$ are, respectively, the
differences of the longitudinal and transverse polarisabilities
between upper and lower quantum state. The polarizabilities of the
hyperfine states of $\HDp$\lyxdeleted{schiller}{Sat Oct 26
17:29:06 2013}{ } have been calculated in Ref.~\cite{hfi} in
absence of electric quadrupole interaction and for zero magnetic
field $B$, employing the Born-Oppenheimer approximation. A
summation method was used, where excited electronic states were
negelected. The method is also applicable if the magnetic field is
finite. The polarizabilities $\alpha$ of the hyperfine states
typically lie in the range of 1 to 100 atomic units, except for
$(v,\, L=0)$ levels, where they are 400 atomic units or larger.
The work put into evidence the strong variation of the
polarisability between different hyperfine state belonging to the
same rovibrational level. The hyperfine-state dependence arises in
the difference\lyxadded{schiller}{Sun Oct 27 01:16:13 2013}{}
$\Delta\alpha^{(l)}-\Delta\alpha^{(t)}$, while the
combination\lyxadded{schiller}{Sun Oct 27 01:16:18
2013}{}$\Delta\alpha^{(l)}+2\Delta\alpha^{(t)}$ is independent of
the upper and lower hyperfine states. The (normalized)
hyperfine-state dependence is precisely obtained from the
summation method, but the magnitudes of the two
\lyxdeleted{schiller}{Sat Oct 26 17:30:26
2013}{quantities}\lyxadded{schiller}{Sat Oct 26 17:30:29
2013}{polarisabilities} are only accurate at a level of a
few~at.u.\lyxadded{schiller}{Sat Oct 26 17:29:41 2013}{
\cite{Comment on accuracy of Stark shift values-1}}. A more
accurate calculation is described in \cite{Korobov and Bekbaev-1},
based on precise variational wavefunctions, which include the
contribution of excited electronic levels. We use the results of
this latter calculation here, which are reported in the tables
above. The values from the two calculation approaches differ by an
amount that scales with the change in vibrational quantum number
and reaches several atomic units for the transitions with
$v=0\rightarrow v'=4,$

According to the tables, many transitions exhibit
\lyxadded{schiller}{Sun Oct 06 17:46:29 2013}{a }differential
polarisability on the order 10~at.u.\lyxadded{Schiller}{Thu Sep 12
21:43:37 2013}{,} which corresponds to a frequency shift
coefficient\lyxdeleted{Schiller}{Thu Sep 12 21:20:39 2013}{ }
$\Delta E_{S}/\langle
E^{2}\rangle=$1.2~mHz/(V/cm)\textsuperscript{2}. We may compare
this with the coefficients of atomic ions used in ion clocks. For
example, it is 0.14 mHz/(V/cm)$^{2}$ for Al\textsuperscript{+} and
1 mHz/(V/cm)$^{2}$ for the octupole transition in
\textsuperscript{171}Yb\textsuperscript{+}. For the latter ion the
associated relative frequency uncertainty in current
state-of-.the-art clocks is at the level of less than $10^{-17}$,
i.e. less than 10~mHz absolute \cite{Huntemann-1}. We assume for
the following that it should be possible to reach a similar
absolute level, 10~mHz, also for $\HDp$, if the transitions have
a\lyxdeleted{Schiller}{Thu Sep 12 21:20:39 2013}{ } polarisability
of 10 at.u.; and correspondingly more if the polarisability is
higher.

\subsection{Potential of promising transitions}

For the rotational and radiofrequency transitions the relative uncertainties
orginating from the Stark shift will generally be larger than for
the rovibrational transitions due to the smaller transition frequencies.

For the radiofrequency transitions, the differential
polarisabilities vanish for $L=0$ levels, since for these levels,
the state polarisabilities are equal for all hyperfine states. The
other transitions considered in Table~\ref{table:RF transitions
shifts} have small or moderate differential polarisabilities. The
947.6~MHz radiofrequency transition in $(1,\,1)$ considered in
Sec.~\ref{sub:Radio-frequency-transitions} exhibits the
differential polarizabilities \lyxadded{schiller}{Sun Oct 27
01:16:04 2013}{}$\Delta\alpha^{(t)}\simeq9\,{\rm
at.u.},\,\Delta\alpha^{(l)}\simeq-17\,{\rm at.u.}$ Following the
argument given in the previous paragraph, the
corr\lyxadded{schiller}{Sat Oct 26 14:02:43 2013}{e}sponding
uncertainty should be controllable at the 0.015~Hz level. The
electric quadrupole shift should be determinable to about the same
level, see Sec.~\ref{sub:Quadrupole-shift-measurement} and the
Zeeman shift inaccuracy was estimated at 0.03~Hz. Thus, in this
particular radiofrequency transition the combined Zeeman,
quadrupole, and Stark systematic shift should be controllable to
approximately 0.05~Hz uncertainty, or $5\times10^{-11}$ in
relative terms.

For the rotational transitions in Table~\ref{tab:rotational
transitions}, we find polarisabilities ranging from intermediate
to large. We estimate that for the stretched-state transition of
$(0,\,0)\rightarrow(0,\,1)$ the total systematic shift uncertainty
could be 0.5~Hz $(5\times10^{-13})$, whereas it could be 0.15~Hz
for the 0.2~MHz component of $(0,\,1)\rightarrow(0,\,2)$, or
$5\times10^{-14}$.

For the one-photon rovibrational transitions, Table~\ref{tab:
rovibrational transitions} contains several with small
differential polarisabilities. For example, the -16 MHz hyperfine
component of $(0,\,3)\rightarrow(3,\,4)$ \lyxadded{schiller}{Sat
Oct 26 13:32:56 2013}{and the -71~MHz hyperfine component of
$(0,\,3)\rightarrow(5,\,4)$, both of which have negligible Zeeman
shift in a 0.02~G magnetic field, }would have approximately
0.01~Hz Stark shift uncertainty, less than the one expected from
the quadrupole shift, 0.1~Hz. The total uncertainy,
$(4-5)\times10^{-16}$, would be dominated by the latter.

Among the stretched-states rovibrational one-photon transitions in
Table~\ref{tab: rovibrational stretched-state transitions} there
are several that have differential polarisabilities of 10~at.u. or
less and therefore contribute much less to the total uncertainty
than the Zeeman effect and the electric quadrupole effect. Here,
too, a total uncertainy of $5\times10^{-16}$ appears possible.

Finally, for the two-photon transitions in Table
\ref{tab:two-photon transitions}, the differential
polarisabilities are moderate to large. For the
$(0,\,0)\rightarrow(2,\,0)$ transition, the Stark effect is the
only nonzero systematic effect of the three types considered. Its
contribution to the transition frequency uncertainty would be
0.14~Hz according to our\lyxdeleted{Schiller}{Thu Sep 12 21:20:39
2013}{ } assumptions, or $1\times10^{-15}$ in relative
terms.\lyxdeleted{Schiller}{Thu Sep 12 21:20:39 2013}{ } A rough
estimate of the light shift is 1~Hz ($1\times10^{-14})$. Thus,
this shift must be measured to the sub-10\% level in order to
reduce the total uncertainty to $1\times10^{-15}$.

\section{Conclusion}

In this paper, we have developed an exact treatment of the interaction
of molecular hydrogen ions with a static electric quadrupole field.
This was simplified by applying the Born-Oppenheimer approximation
and we derived an approximate effective Hamiltonian. We computed the
corresponding coupling coefficients $E_{14}$ for the three nonradioactive
molecular hydrogen ion species. The quadrupole shift can be obtained
with sufficient accuracy by applying first-order perturbation theory.
It is worth noting that the computational scheme outlined here may
be useful in estimating similar effects in the spectroscopy of exotic
bound systems (such as muonic hydrogen molecular ions \cite{shima,dtm})
in the liquid or solid phase. The shift of energy levels with zero
rotational angular momentum vanishes. Experimentally, the quadrupole
shift can be nulled by measuring the mean of the transition frequencies
when the magnetic field is aligned along three orthogonal directions.
This holds true for all molecular hydrogen ions, and is due to the
smallness of the quadupole interaction.

We evaluated the electric quadrupole shifts of a large number of
transitions in $\HDp$, the hydrogen molecular ion most intensively
studied with high-resolution optical spectroscopy to date. We have
considered those radio-frequency, rotational, rovibrational one-
and two-photon E1 transitions that have low, vanishing, or
opposite equal Zeeman shifts, and that are therefore of interest
for precision spectroscopy. The radio-frequency (and rotational)
transitions, for which the relative uncertainty is higher than for
the rovibrational ones, are of interest for a test of the
hyperfine hamiltonian of the molecule, while the the rovibrational
transitions are of interest for QED tests, fundamental constants
metrology and equivalence principle tests.

For the rovibrational transitions we find one-photon transitions
of very low Zeeman shift and two-photon transitions that are free
of Zeeman shift and of quadrupole shift. In the one-photon
transitions of smallest quadrupole shift, it is of relative
magnitude close to $1\times10^{-15}$. However, if the nulling
procedure is applied, the uncertainty in the residual quadrupole
shift can be reduced to this level for essentially all
rovibrational transitions.

Combining these considerations with earlier analyses of the blackbody
shift and a\lyxdeleted{schiller}{Sat Oct 26 20:52:04 2013}{n} \lyxadded{schiller}{Sat Oct 26 20:52:08 2013}{recent
precise }evaluation of the Stark shift\lyxdeleted{schiller}{Sat Oct 26 13:37:41 2013}{
}, we conclude that for a few selected rovibrational transitions of
the $\HDp$ ion a relative frequency inaccuracy at the $5\times10^{-16}$
level should be achievable, under realistic assumptions. This inaccuracy
is limited by the accuracy with which the systematic shifts can be
determined, which is ultimately limited by the statistical uncertainty
of measuring the transition frequ\lyxadded{schiller}{Sat Oct 26 17:43:30 2013}{e}n\lyxdeleted{schiller}{Sat Oct 26 17:43:31 2013}{e}cies\lyxdeleted{schiller}{Sat Oct 26 17:43:27 2013}{.
}. Because of the relatively short lifetimes of the vibrational levels,\lyxdeleted{Schiller}{Thu Sep 12 21:20:39 2013}{
} correspondingly long integration times are therefore necessary to
reduce the statistical uncertainty to the above level.

We have not computed the shifts of individual energy levels of $\Htwop$
and $\Dtwop$ in this paper. However, the numerical similarity of
their coefficients $E_{14}$ to those of $\HDp$ indicates that their
quadrupole shifts will be similarly small. In this context, their
distinctive feature is the extremely small natural linewidth of their
transitions and the different spin structure of the transitions. Because
of the smaller linewidth of the homonuclear ions, the statistical
inaccuracy can in principle be significantly lower than in $\HDp$.
Thus, their potential for a molecular ion clock should be investigated
in future studies.

\section{Acknowledgments}

This work has been supported by grant SCHI 431/19-1 of the
Deutsche Forschungsgemeinschaft. We are grateful to V.I. Korobov
and A. Bekbaev for making available to us their results on the
polarisabilities and of the hyperfine hamiltonian coefficients.

\end{document}